\documentclass[amsmat,amssymb,amsfonts,aps,pra,reprint,twocolumn,superscriptaddress,floatfix,longbibliography]{revtex4-2}

\usepackage{graphicx}
\DeclareGraphicsExtensions{.eps}

\usepackage{xcolor}
\usepackage{amsmath}
\usepackage{blindtext, rotating}
\usepackage{braket}
\usepackage{multirow}
\usepackage{tabularx}
\usepackage{dsfont}
\usepackage{booktabs, siunitx}
\usepackage{hyperref}
\usepackage{placeins}
\usepackage{stmaryrd}

\begin{document}

\title{Testing complementarity on a transmon quantum processor}

\author{Pedro M. Q. Cruz}
\email{pedro.cruz@inl.int}
\affiliation{QuantaLab, International Iberian Nanotechnology Laboratory (INL), 4715-330 Braga, Portugal}
\affiliation{Faculdade de Ci\^encias da Universidade do Porto, 4169-007 Porto, Portugal}

\author{J. Fern\'{a}ndez-Rossier}
\altaffiliation[On leave from ]{Departamento de F\'{i}sica Aplicada, Universidad de Alicante, 03690 San Vicente del Raspeig, Spain.}
\affiliation{QuantaLab, International Iberian Nanotechnology Laboratory (INL), 4715-330 Braga, Portugal}

\date{\today}

\begin{abstract}
We propose quantum circuits to test interferometric complementarity using symmetric two-way interferometers coupled to a which-path detector. First, we consider the two-qubit setup in which the controlled transfer of path information to the detector subsystem depletes interference on the probed subspace, testing the visibility-distinguishability tradeoff via minimum-error state discrimination measurements. Next, we consider the quantum eraser setup, in which reading out path information in the right basis recovers an interference pattern. These experiments are then carried out in an IBM superconducting transmon processor. A detailed analysis of the results is provided. Despite finding good agreement with theory at a coarse level, we also identify small but persistent systematic deviations preventing the observation of full particlelike and wavelike statistics. We understand them by carefully modeling two-qubit gates, showing that even small coherent errors in their implementation preclude the observation of Bohr's strong formulation of complementarity.
\end{abstract}

\maketitle

\section{Introduction}

Complementarity is one of the most intriguing aspects of quantum mechanics, as it sets limits to our capability to gather simultaneous information about different observables. It was first discussed in the context of wave-particle duality by the founders of the theory, but soon became clear that it applies to any set of noncommuting observables. The central conundrum of complementarity is often formulated in terms of a quantum object that can take two alternatives, or paths, that we denote by $|0\rangle$ and $|1\rangle$. The wave function is prepared in an equal superposition of the paths with a relative phase added:
\begin{equation}
|\psi\rangle= c_0 |0\rangle + c_1 e^{i\phi} |1\rangle.
\label{eq:two-paths-phase}
\end{equation}
As the phase is changed and the expectation value of a complementary observable is measured, an interference pattern emerges with an amplitude proportional to $\operatorname{Re}{ \left\{ c_0^*c_1 \right\} }$. This can only occur if the object evolves through both ways simultaneously, interfering with itself (wavelike character). If a system is set up to gather information about which path the object actually realizes (particlelike character), the interference contrast, or visibility, is washed out.

This modern understanding of the interplay between which-path information and interference visibility began its formulation in 1927 with Bohr's \emph{principle of complementarity} \cite{bohr1928quantum}. Building on previous achievements \cite{debroglie1924recherches, heisenberg1927uber}, he first argued complementarity to be a fundamental quantum feature whereby objects possess pairs of properties the knowledge of which is mutually exclusive, the observation of either unambiguous particle trajectories or interference fringes being just one example.

In the same year, Einstein conceptualized a recoiling double-slit interferometer hoping to falsify Bohr's principle \cite{bohr1949discussion}. This first which-path thought experiment was discussed and followed by other proposals over the years \cite{heisenberg1930physical, feynman1965feynman, wheeler1978past}, but only analyzed in depth half a century later \cite{wootters1979complementarity}. On that analysis of the momentum exchange between the photon and the slit, the tradeoff between the interference visibility and the amount of which-path information retrieved was quantified. This description of the partial knowledge obtained from complementary observables confirmed and extended Bohr's original dichotomous version in which complete knowledge of one of them implies maximal uncertainty about the other.

Soon after, other which-path experiments were proposed to test complementarity enforced by physical mechanisms not arising from the position-momentum uncertainty relation \cite{scully1989spin, scully1989quantum}. The information-theoretic treatment initiated in \cite{wootters1979complementarity} was further pursued by others \cite{deutsch1983uncertainty, mittelstaedt1987unsharp, greenberger1988simultaneous} and a \emph{duality relation} for general bipartite two-dimensional systems was obtained by Jaeger \emph{et al.} \cite{jaeger1995two} and Englert \citep{englert1996fringe}, connecting the visibility ${\cal V}$ (amplitude) of the pattern produced by the interfering object to the amount of which-path information ${\cal D}$ (see Eq.~\ref{eq:D-definition}) captured by the detector:
\begin{equation}
{\cal V}^2+{\cal D}^2\leq1.
\label{eq:VisDis}
\end{equation}

This analysis was then extended to the context of imperfect detection and quantum erasure \cite{bjork1998complementarity}, as well as more than two interferometric paths \cite{durr2001quantitative}. After the turn of the century, it was also shown that the \emph{a posteriori} distinguishability of the paths, ${\cal D}$, contains two parts to it: the \emph{a priori} predictability and the genuine quantum correlations, as quantified by the concurrence entanglement measure. Eq.~(\ref{eq:VisDis}) was alternatively cast as a triality and generalized to bipartite systems of any dimension \cite{jakob2007complementarity, jakob2010quantitative}.

Originally, Englert argued that the duality relation in Eq.~(\ref{eq:VisDis}) was independent of Heisenberg's uncertainty principle. Since then, however, a new entropic formulation of the latter has been discovered, allowing its equivalence to the former to be established for multipath interferometers \cite{coles2014equivalence, coles2016entropic}. Other wave-particle duality relations based on different information quantifiers and applicable to Hilbert spaces of arbitrary dimension have also recently been introduced \cite{angelo2015wave, bagan2016relations}.

Complementarity, and the controlled depletion of interference by including which-path detectors in the experimental setup, have been observed in very ingenious experiments probing diverse interferometer-detector systems, including atom-photon \cite{durr1998origin, bertet2001complementarity}, photon-photon \cite{herzog1995complementarity, kim2000delayed, walborn2002doubleslit, neves2009control, peruzzo2012aquantum, kaiser2012entanglement}, electron-electron \cite{buks1998dephasing, sprinzak2000controlled, neder2007entanglement}, coupled atomic nuclei in molecules \cite{teklemariam2001NMR, teklemariam2002quantum, peng2003interferometric}, and superconducting circuits \cite{zheng2015quantum, liu2017atwofold, bienfait2020quantum}. Wave-particle duality has also been observed in optical interferometers in which the path information is not stored in a sub-system but carried by an additional degree of freedom of the photons along their path \cite{kwiat1992observation, schwindt1999quantitative}.

Today, with the advent of transmon quantum computers and IBM Quantum (IBM~Q) cloud services \cite{ibmq}, new tests of complementarity are being performed. This platform has been used to carry out which-path experiments with multipath interferometers and extract quantifiers of visibility and distinguishability using an unambiguous state discrimination detector readout in \cite{amico2020simulation}. The triality relation in \cite{jakob2010quantitative} was tested on two qubits with a tomography-based circuit in \cite{schwaller2021evidence}. Moreover, a state tomography approach was also used in \cite{pozzobom2021experimental} to test different complementarity relations obtained from density matrix properties of random states of up to three qubits.

Here, we study two classes of which-path experiments using simple two-qubit circuits to implement a symmetric two-way interferometer coupled to a single detector. First, we design a setup in which the controlled transfer of path information to the detector subsystem depletes interference on the probed subspace, testing Eq.~(\ref{eq:VisDis}) by quantifying distinguishability via minimum-error state discrimination measurements. Then, we consider a nondelayed quantum-eraser configuration in which the detector readout basis destroys path information allowing interference to be recovered. We experiment on a superconducting quantum processor made available online by IBM~Q. In this architecture, microwave pulses drive arbitrary single-qubit rotations and entanglement of fixed-frequency transmons. We pay close attention to fine details in the experimental data to accurately interpret results and draw their fundamental consequences.

The rest of this paper is organized as follows. In section \ref{sec:wp-theo}, we introduce the abstract quantum circuits that implement optimal which-path detection and quantum erasure on any gate based quantum computer. In section \ref{sec:exp}, we present the experimental platform and the physical circuits, describe the methodology, and analyze experimental results. In section \ref{sec:RCM}, reassessment of the physical operations and model evaluation are performed to understand the encountered deviations from the theory. Finally, section \ref{sec:disc} discusses the main takeaways and presents our conclusions.

\section{Which-path quantum circuits \label{sec:wp-theo}}

\begin{figure}
\includegraphics[width=0.83\linewidth]{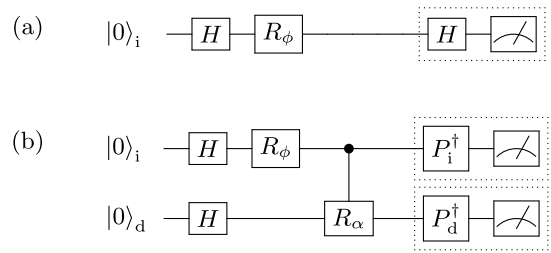}
\caption{(a) Single qubit circuit equivalent to a Mach-Zehnder interferometer. (b) Same setup, coupling a second qubit, $\mathrm{q_d}$, to the interferometer qubit, $\mathrm{q_i}$, to act as a which-path detector. The readout basis is chosen by defining $P_\mathrm{i}^\dagger$ and $P_\mathrm{d}^\dagger$ before the $Z$ basis measurement. Setting $P_\mathrm{i}^\dagger=H$ and varying the value of $\alpha$, interference on $\mathrm{q_i}$ can be depleted, independently of reading out $\mathrm{q_d}$. Setting $P_\mathrm{d}^\dagger=O_\alpha^\dagger$ maximizes which-path information retrieval. Making $P_\mathrm{d}^\dagger=\mathds{1}$ instead erases path information rendering detection outcomes useless while allowing full-visibility interference patterns on $\mathrm{q_i}$ to be recovered. Both measurements occur simultaneously. A delayed-choice experiment would probe the detector after the interferometer.}
\label{fig:whichpath}
\end{figure}

Our starting point is the quantum circuit representation of a Mach-Zehnder interferometer (MZI) \cite{cleve1998quantum} (see Fig.~\ref{fig:whichpath}a). The first Hadamard gate prepares an equal superposition of the two qubit states, $\frac{1}{\sqrt{2}}\left(|0\rangle+|1\rangle\right)$, very much like the first beam splitter provides two alternatives in the MZI. The phase gate $R_{\phi}$ introduces a relative phase between the two alternatives, yielding the state in Eq.~(\ref{eq:two-paths-phase}) with $c_0=c_1=\frac{1}{\sqrt{2}}$. The second $H$ gate, equivalent to the second beam splitter in the MZI, is used to measure the $X$ operator with a readout in the $Z$ basis. The output state $\left| \psi' \right\rangle = H |\psi\rangle$ on $\mathrm{q_i}$ satisfies
\begin{equation}
\langle \psi'| Z|\psi'\rangle = \langle \psi| H^{\dagger} Z H|\psi\rangle = \langle \psi|X|\psi\rangle = \cos\phi.
\label{eq:1qb}
\end{equation}
Thus, by measuring $\ket{\psi'}$ in the computational basis we obtain the expectation $\langle \psi|X|\psi\rangle$ and detect interference, probing thereby the linear superposition and the added relative phase. In this setup, we gain no information about which path, i.e. state $|0\rangle$ or $|1\rangle$, was taken by the qubit before the $X$ basis measurement.

The which-path detector is implemented by adding a second qubit, $\mathrm{q_d}$, to the circuit (Fig.~\ref{fig:whichpath}b). The controlled phase gate introduces a relative phase $\alpha$ in the joint interferometer-detector state when both qubits are in state $\ket{1}$. The wave function of the two-qubit system after this operation, in the basis $\left|xy\right\rangle = \left|x\right\rangle_\mathrm{i} \otimes \left|y\right\rangle_\mathrm{d}$, is
\begin{equation}
\ket{\Psi} = \frac{1}{\sqrt{2}}\left(\ket{0}\otimes \ket{\delta_0} + e^{i\phi} \ket{1} \otimes \ket{\delta_1} \right),
\label{eq:2q}
\end{equation}
where $\ket{\delta_{x}}$, $x\in\left\{ 0,1\right\}$, denotes the state of the detector qubit conditioned to the path $\ket{x}$ taken by the first one,
\begin{eqnarray}
\ket{\delta_0} &=& \frac{1}{\sqrt{2}}\left(\ket{0} + \ket{1}\right), \nonumber \\
\ket{\delta_1} &=& \frac{1}{\sqrt{2}}\left(\ket{0} + e^{i\alpha} \ket{1}\right).
\label{eq:detector-states}
\end{eqnarray}
The {\em distinguishability} ${\cal D}$ of these states is quantified by the trace distance of their density matrices, yielding
\begin{align}
{\cal D} & \equiv \frac{1}{2}\left\Vert \ket{\delta_0}\bra{\delta_0} - \ket{\delta_1}\bra{\delta_1} \right\Vert _{T} \label{eq:D-definition} \\
 & = \sqrt{1 - |\langle\delta_0|\delta_1\rangle|^2} = \sqrt{\sin^2\frac{\alpha}{2}}, \label{eq:D-purestate}
\end{align}
where we have used 
\begin{equation}
\langle\delta_0|\delta_1\rangle = \frac{1}{2}\left(1+e^{i\alpha}\right).
\label{eq:over}
\end{equation}
Thus, ${\cal D}$ is maximal for $\alpha=\pi$ and null for $\alpha=0$.

Since we want to measure the detector to correctly identify the path taken by the first qubit, we should maximize the probability of success. This is accomplished by obtaining the maximum which-path information leaked with a measurement in the optimal basis \cite{helstrom1976quantum}. For that, we use the eigenvectors of $\ket{\delta_0}\bra{\delta_0}-\ket{\delta_1}\bra{\delta_1}$ to build an operator $O_{\alpha}$, which we conjugate transpose to obtain
\begin{equation}
O_{\alpha}^\dagger=\frac{1}{\sqrt{2}}\left(\begin{array}{cc}
1 & ie^{-\frac{i\alpha}{2}}\\
1 & -ie^{-\frac{i\alpha}{2}}
\end{array}\right).
\label{eq:optimal}
\end{equation}
Making $P_\mathrm{d}^\dagger=O_\alpha^\dagger$ in the circuit of Fig.~\ref{fig:whichpath}b transforms the reduced density matrix over $\mathrm{q_d}$ to $\hat{\rho}_x = O_\alpha \ket{\delta_x} \bra{\delta_x} O^\dagger_\alpha$, which measured in the computational basis yields the outcome $y$ on the detector with maximal probability of successfully identifying the correct path $\ket{x}$ on $\mathrm{q_i}$:
\begin{equation}
\mathrm{p}_{\mathrm{succ}} = \frac{1}{2} \left( \mathrm{p} \left(0_{\mathrm{d}}\middle|\hat{\rho}_0\right) + \mathrm{p} \left(1_{\mathrm{d}}\middle|\hat{\rho}_1\right) \right) = \frac{1}{2} \left( 1 + {\cal D} \right).
\label{eq:p_success}
\end{equation}
The significance of the distinguishability is therefore that it bounds the maximum possible success probability. For $\alpha=\pi$, $\hat{\rho}_0=\mathds{1}-\hat{\rho}_1 = \ket{0}\bra{0}$, allowing for fully determined path information. If we run the circuit in Fig.~\ref{fig:whichpath}b with $P_\mathrm{i}^\dagger=\mathds{1}$ and $P_\mathrm{d}^\dagger=O_\alpha^\dagger$ we can obtain ${\cal D}$ empirically from
\begin{equation}
\mathrm{p}_{\mathrm{succ}} = \frac{1}{2}\mathrm{p}\left(0_\mathrm{d}\middle|0_\mathrm{i}\right) + \frac{1}{2}\mathrm{p}\left(1_\mathrm{d}\middle|1_\mathrm{i}\right),
\label{eq:p_success_empirical}
\end{equation}
where $\mathrm{p}\left(y\middle|x\right)=\mathrm{p}(xy)/\sum_{y} \mathrm{p}(xy)$ stands for the conditional probability and $\mathrm{p}(xy)$ is the measurement probability of the $\ket{xy}$ output state.

\subsection{Which-path: interference vs. detection}

The interference pattern produced by the first qubit in the circuit of Fig.~\ref{fig:whichpath}b with $P_\mathrm{i}^\dagger=H$ is obtained from
\begin{equation}
\langle X \rangle \equiv \operatorname{Tr}{ \hat{\sigma} X\ },
\end{equation}
where $ \hat{\sigma}$ is the reduced density matrix on $\mathrm{q_i}$ before the basis change operation, associated to the two-qubit state from Eq.~(\ref{eq:2q}) tracing over the states of $\mathrm{q_d}$,
\begin{equation}
\hat{\sigma}\equiv \operatorname{Tr}_{\mathrm{q_d}}{ |\Psi\rangle\langle\Psi| }.
\end{equation}
After straightforward algebra, we obtain
\begin{equation}
\langle X \rangle = \operatorname{Re}{ \left\{ e^{i\phi} \langle\delta_0|\delta_1\rangle\right\} }= \frac{\cos\phi + \cos(\phi+\alpha)}{2}.
\label{eq:avX}
\end{equation}
Thus, it is apparent that the contrast of the interference pattern along $\phi$ is controlled by the overlap of the possible states of the detector qubit. When this overlap, and thereby contrast, is maximal ($\alpha=0$), which-path information is null. On the other hand, for $\alpha=\pi$ there is no interference and maximal which-path information is stored on the detector, allowing for its readout to unambiguously identify the path taken by the first qubit. In other words, it is by maximally entangling the two-qubit system ($\alpha=\pi$) that we get $\mathrm{q_i}$ to display classical particlelike statistics with no interference visibility, regardless of reading out the detector.
\begin{figure}[t]
\includegraphics[width=1\linewidth]{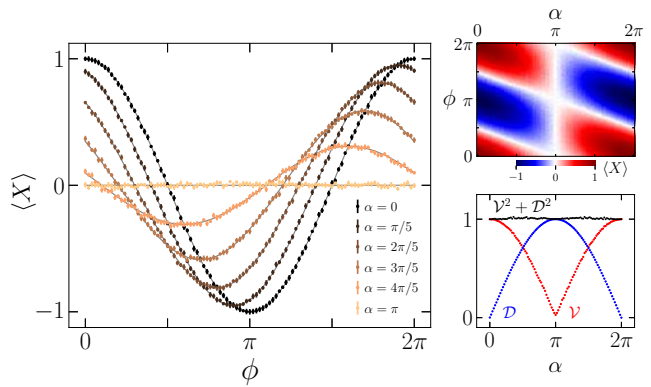}
\caption{In-silico pure-state simulations obtained with two families of circuits of the type of Fig.~\ref{fig:whichpath}b, executed with $8192$ shots for each of 10201 pairs of values $\alpha, \phi \in [0,2\pi]$. The first family consists of the circuit with $P_\mathrm{i}^\dagger=H$ and $P_\mathrm{d}^\dagger=O_\alpha^\dagger$. The top right panel shows the measured $\langle X \rangle$ on $\mathrm{q_i}$ for all these circuits. The left side panel illustrates a subset of the data for six different values of $\alpha$: actual results (dots with $\pm2\sigma_{\left\langle X\right\rangle }$ errorbar) are compared against the theoretical expectation given by Eq.~(\ref{eq:avX}) (solid lines). The third panel shows the results for visibility and distinguishability; data in blue are obtained from the second family of circuits, where the circuit in Fig.~\ref{fig:whichpath}b is executed with $P_\mathrm{i}^\dagger=\mathds{1}$ and $P_\mathrm{d}^\dagger=O_\alpha^\dagger$, and Eqs.~(\ref{eq:p_success}\,--\,\ref{eq:p_success_empirical}) are employed to extract ${\cal D}$.}
\label{fig:B-circ_insilico}
\end{figure}

Experimentally, this interference pattern is obtained from $\langle X \rangle = \mathrm{p}(0_\mathrm{i})-\mathrm{p}(1_\mathrm{i})$, approximating the output probabilities with the relative frequencies of the readouts. This process is subject to shot noise, the standard deviation of which scales with $\frac{1}{\sqrt{S}}$, where $S$ is the number of readouts, or shots. In Fig.~\ref{fig:B-circ_insilico} we show the results of an in-silico simulation generated by ramping the phase $\phi$ in the first qubit, for several values of the control parameter $\alpha$. With $8192$ shots per circuit, the curves traced by the data are in perfect agreement with Eq.~(\ref{eq:avX}), and the 95.4\% confidence interval contained by $\left\langle X\right\rangle\pm2\sigma_{\left\langle X\right\rangle}$ is almost imperceptible. As we vary $\alpha$ from zero to $\pi$, the amplitude of the interference pattern gradually vanishes.

To summarize these observations we define the interference visibility and compute it with Eq.~(\ref{eq:avX}):
\begin{equation}
{\cal V} \equiv \frac{\langle X \rangle_{\alpha}^{\max}(\phi)-\langle X \rangle_{\alpha}^{\min}(\phi)}{2+\langle X \rangle_{\alpha}^{\max}(\phi)+\langle X \rangle_{\alpha}^{\min}(\phi)} = \sqrt{\cos^2 \left(\frac{\alpha }{2}\right)}.
\label{eq:V-purestate}
\end{equation}
Using Eq.~(\ref{eq:over}), visibility is related to the overlap
\begin{equation}
{\cal V} = |\langle\delta_0|\delta_1\rangle|,
\end{equation}
and therefore obeys the well-known \cite{jaeger1995two, englert1996fringe} result
\begin{equation}
{\cal V}^2+{\cal D}^2 = 1.
\label{eq:englert-equality}
\end{equation}
This provides a mathematical description of the complementarity relation between distinguishability and visibility: imprinting which-path information of the first qubit into the second one gradually destroys interference. In the next section, we address the question of whether the interference is actually destroyed or if we could recover it by interrogating the detector differently.

\subsection{Quantum erasure}

As we have seen in the previous section, the interference pattern of the first qubit is compromised if the which-path information is stored in the detector, even if this second qubit is not measured. We now consider the setup of Fig.~\ref{fig:whichpath}b when the state of the detector is instead measured with $P_\mathrm{d}^\dagger=\mathds{1}$, keeping $P_\mathrm{i}^\dagger=H$.

This detector readout destroys the which-path information by collapsing the state directly into the $Z$ basis. Given that both $\ket{0}$ and $\ket{1}$ have equal weight on that projection, path information is lost regardless of the value of $\alpha$. To see how the irreversible destruction of the which-path information can restore interference, we write the quantum state just before readout as
\begin{align}
\ket{\Psi_\mathrm{out}} & = 
\frac{1+e^{i\phi}}{2\sqrt{2}}  \ket{00} + 
\frac{1+e^{i\phi+i\alpha}}{2\sqrt{2}} \ket{01} \nonumber \\
& \;
+ \frac{1-e^{i\phi}}{2\sqrt{2}} \ket{10} 
+  \frac{1-e^{i\phi+i\alpha}}{2\sqrt{2}}\ket{11}.
\label{eq:out}
\end{align}
With that, we have the following probabilities for all the possible measurement outcomes:
\begin{align}
\mathrm{p}(00) & = \frac{1}{4}(1+\cos\phi), \nonumber \\
\mathrm{p}(01) & = \frac{1}{4}(1+\cos(\phi+\alpha)), \nonumber \\
\mathrm{p}(10) & = \frac{1}{4}(1-\cos\phi) \nonumber, \\
\mathrm{p}(11) & = \frac{1}{4}(1-\cos(\phi+\alpha)),
\end{align}
where $\mathrm{p}(xy)= \left|\left\langle \Psi_\mathrm{out}|xy\right\rangle \right|^{2}$. Let us then look at the expectation value of the interferometer qubit in the $X$ basis, broken-down according to the outcome of the second qubit, which will be $\ket{0}$ or $\ket{1}$ with equal probability. Conditioning on these events we get
\begin{align}
\langle X_0 \rangle & \equiv \mathrm{p}\left(0_\mathrm{i}\middle|0_\mathrm{d}\right) - \mathrm{p}\left(1_\mathrm{i}\middle|0_\mathrm{d}\right) = \cos\phi, \label{eq:X0conditioned} \\
\nonumber \\
\langle X_1 \rangle & \equiv \mathrm{p}\left(0_\mathrm{i}\middle|1_\mathrm{d}\right) - \mathrm{p}\left(1_\mathrm{i}\middle|1_\mathrm{d}\right) = \cos(\phi+\alpha). \label{eq:X1conditioned}
\end{align}

It is clear that by discriminating $\mathrm{q_i}$ readouts following the guidance of the detector, two $\mathcal{V}=1$ interference patterns are obtained, shifted by $\alpha$. It is only when these measurements are aggregated, disregarding the detector output and protecting thereby path information, that we return to the situation where the interference visibility is compromised by $\alpha$, recovering the result in Eq.~(\ref{eq:avX}):
\begin{equation}
\frac{1}{2} \langle X_0 \rangle + \frac{1}{2} \langle X_1 \rangle = \langle X \rangle.
\end{equation}

Note that if we perform the same break-down procedure with $P_\mathrm{d}^\dagger=O_\alpha^\dagger$ instead we cannot recover full-visibility interference patterns like these. Nature seems to know that our measurement projection has already obtained some which-path information, therefore it does not let us observe the complete wavelike behavior of the system. Similar approaches to quantum erasure have been described in \cite{scully1991quantum, bertet2001complementarity, ionicioiu2011proposal}.

\section{Experimental results \label{sec:exp}}

We now discuss the experimental testing of the circuits in the previous section. For that matter, we took advantage of the IBM~Q platform \cite{ibmq}, which provides remote access to digital quantum processors based on superconducting fixed-frequency transmon qubits coupled via the cross-resonance gate. In particular, we used the Qiskit open-source software package \cite{qiskit} to prepare several experiments for execution in \emph{ibmq\_toronto}, a 27-qubit \emph{Falcon r4} processor with a heavy-hexagon qubit lattice \cite{chamberland2020topological} achieving \emph{quantum volume} 32.

The core of a transmon is formed by two superconducting islands coupled through two Josephson junctions (JJs). The relevant dynamical quantities of the JJs are the electron imbalance across the junction and the difference of the phases of the superconducting order parameter, which for all practical purposes can be considered conjugate variables. The qubit states $\ket{0}$ and $|1\rangle$ are encoded in the two lowest energy states of the Hamiltonian describing this system. The control and readout of the JJs is implemented by means of their capacitive coupling to the electromagnetic modes confined in a transmission line resonator. For details on the working principles of the device we refer the reader to the original papers \cite{blais2007quantum, koch2007charge, majer2007coupling} and recent reviews \cite{krantz2019aquantum, blais2020circuit}.

IBM~Q processors have been previously used to test other fundamental aspects of quantum mechanics, such as Bell and Mermin inequalities \cite{alsina2016experimental, garcia2018five}. Quantum computing systems based on transmons are also being developed by others, such as Rigetti \cite{rigettiweb}, Google Quantum AI \cite{googleqai}, and Quantum Inspire \cite{quantuminspire}.

\subsection{Methodology}

The methodology used to test any of the circuits in Fig.~\ref{fig:whichpath} involved two steps: first, the estimation of the relevant observables from a sample of $S$ shots for each of $C$ values for the variables of the circuit being tested. At a repetition time of 500 $\mu s$ per shot, the total processing time of an experiment adds up to about $500\,CS$ $\mu s$. The set of circuits comprising a full experiment was executed in batches of 896. Four additional measurement error mitigation \cite{chen2019detector, maciejewski2020mitigation, qiskittextbook2020} circuits were included in each batch, except for the experiment with the circuit in Fig.~\ref{fig:whichpath}a. Within each batch, one shot of each circuit was executed sequentially before moving on to each of the remaining shots. Batches ran one after another until completion of all the executions required to carry out an experiment. Each circuit in a circuit family is defined by its $\phi$ and $\alpha$ values. Hence, the order in which circuits are executed sets the order in which this space is probed. Because of the second step of the methodology, this space was probed uniformly at random.

In the second step, the previous $C$ estimates were employed to find a global experimental model $f(\phi,\alpha)$ through a curve-fitting procedure minimizing
\begin{equation}
\chi^2=\sum_{j=1}^{C}\left(\frac{z_{j}-\hat{z}_{j}}{\sigma_{j}}\right)^{2},
\label{eq:chi-squared}
\end{equation}
where $\hat{z}_{j}\equiv f\left((\phi,\alpha)_{j}\right)$, and the standard error $\sigma_j$ quantifies the dispersion associated with each estimate $\hat{z}_j$. Although the measurement error mitigation procedure is used to infer sources of error during data analysis, all models below are fitted to raw, unmitigated data. 

In order to set up the model $f(\phi,\alpha)$ to fit the data, the theoretical expression $g(\phi,\alpha)$ obtained for the case of ideal unitary execution of a given quantum circuit is modified with two additional features. First, one fit parameter is introduced per circuit variable, replacing
\begin{equation}
\begin{array}{c}
\phi\rightarrow\phi+\Theta_{1}, \\
\alpha\rightarrow\alpha+\Theta_{2},
\end{array}
\label{eq:bigthetas}
\end{equation}
in the $g(\phi,\alpha)$ model to account for calibration errors in the physical gates. These two additional parameters improve the modeling of the experimental data and their seemingly \emph{ad hock} introduction will be justified in section \ref{sec:RCM}. Second, since the quantum processor is not perfectly isolated from the environment, the dynamics over this subspace is nonunitary. To accommodate for the existence of a classical mixture with identically distributed noise on all the output distributions from an experiment, the pure state expression is scaled by $\eta$ and a bias parameter $\varepsilon$ is introduced. The final experimental model carries at most four parameters and is written as
\begin{equation}
f(\phi,\alpha)=\eta\;g(\phi,\alpha)+\varepsilon.
\label{eq:experimental-model}
\end{equation}

\begin{figure}
\includegraphics[width=0.9\linewidth]{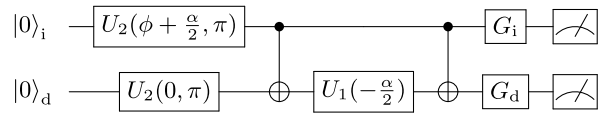}
\caption{Actual physical circuit implemented to reproduce Fig.~\ref{fig:whichpath}b in terms of the elementary gateset of the device. Final gates are defined for each readout basis according to the correspondence: $P_\mathrm{i}^\dagger=H\Rightarrow G_\mathrm{i}=U_2(0,\pi)$, $P_\mathrm{i}^\dagger=\mathds{1}\Rightarrow G_\mathrm{i}=\mathds{1}$, $P_\mathrm{d}^\dagger=O^\dagger_\alpha\Rightarrow G_\mathrm{d}=U_2(0, \frac{3\pi}{2})$, and $P_\mathrm{d}^\dagger=\mathds{1}\Rightarrow G_\mathrm{d}=U_1(\alpha/2)$.}
\label{fig:transp-circs}
\end{figure}

After completion of the curve-fitting procedure, the autocorrelation of the residuals is checked in the order of the execution of the circuits. If observed, it is associated with drifts in the calibration and coherence properties of the device during the experiment. The attribution is made possible by the execution protocol whereby the $(\phi, \alpha)$ space is sampled at random, since it removes the correlation component coming from the possible use of an incorrect model to fit the data. In order to detect autocorrelation without making any strong assumptions about the generating error process, the residuals are divided into positive or negative at their average value and the nonparametric single-sample Wald-Wolfowitz runs test \cite{wald1940test} is performed on the sequence. Evidence for autocorrelation is found if the test rejects the null hypothesis for errors to be independent and identically distributed.

All required circuits in an experiment must be executed on the same two-qubit pair and within a single calibration round of the system to maintain calibration errors consistency across the tested variable space. For this reason, the random sampling protocol brings an additional benefit. If the calibration of the device changes before sampling all the intended variable values, it is still possible to fit the model with the already obtained values uniformly spread over the variable space. Furthermore, if mild environmental drifts in the coherence properties do occur, randomly sampling $(\phi, \alpha)$ allows filtering outliers uniformly in this space.

At the time of the experiment, IBM~Q hardware uses the controlled-\textsc{not} (\textsc{cnot}) gate together with a fixed set of parametrized single-qubit gates to implement any quantum circuit. For this reason, the circuits described in Fig.~\ref{fig:whichpath} were converted to the native operations as represented in Fig.~\ref{fig:transp-circs}. Only two single-qubit physical gates were used:
\begin{align}
U_{1}\left(y\right) & = \left(\begin{array}{cc}
1 & 0\\
0 & e^{iy}
\end{array}\right), \nonumber \\
U_{2}\left(x,y\right) & = \frac{1}{\sqrt{2}}\left(\begin{array}{cc}
1 & -e^{iy}\\
e^{ix} & e^{i(x+y)}
\end{array}\right).
\label{eq:ibm-base-ops}
\end{align}
One might notice this circuit can be simplified for some values of $\phi$ and $\alpha$. However, such simplifications should not be done experimentally since we must ensure all the results are obtained from the same number of operations and execution time. This guarantees adherence of the whole dataset to the same experimental model.

Finally, the distribution of residuals is also inspected in the $(\phi, \alpha)$ space to assess any pattern indicating the use of an incorrect model to fit the data. Model adequacy issues are discussed further in section \ref{sec:RCM}.

\subsection{Two-way interferometer}

\begin{figure}
\includegraphics[width=0.8\linewidth]{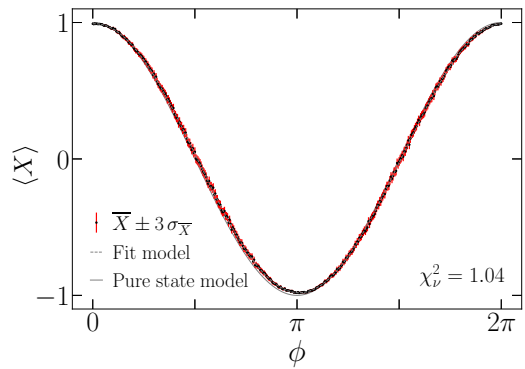}
\caption{Experimental results for the circuit in Fig.\ref{fig:whichpath}a. The $\overline{X}$ estimations are represented with almost imperceptible $\pm3\sigma_{\overline{X}}$ confidence intervals and adjust very well to the properly constructed Eq.~(\ref{eq:experimental-model}) model, as measured by $\chi_\nu^2$. This experimental model is also very close to the pure state model, Eq.~(\ref{eq:1qb}).}
\label{fig:A-circ_ibmq}
\end{figure}

Our first experimental results were obtained for the setup in Fig.~\ref{fig:whichpath}a, taking $S=8192$ shots for each of $C=401$ evenly spaced values of $\phi\in\left[0,2\pi\right]$ (see Fig.~\ref{fig:A-circ_ibmq}). This one-qubit circuit is used as a control experiment to confirm that an interference pattern is observed. For each $\phi$, we registered the readout $X_j$, collecting then all readouts to compute the relative frequencies ($\overline{\mathrm{p}}$) of the two possible outcomes and estimate $\langle X \rangle$ from the mean $\overline{X} = \overline{\mathrm{p}}(0)-\overline{\mathrm{p}}(1)$ with standard error $\sigma_{\overline{X}}=s/\sqrt{S}$, where $s^{2}=\sum_{j=1}^{S}\frac{\left(X_j-\overline{X} \right)^{2}}{S-1}$ is the sample variance.

Based on Eq.~(\ref{eq:1qb}), these data were modeled with an experimental model $\overline{X}=\eta\cos(\phi+\Theta_1)+\varepsilon$. For the chosen $S$ and $C$, the minimization procedure finds very good agreement between the data and the model, scoring $\chi_\nu^2\equiv\chi^2/\mathrm{dof}\approx1.04$, where $\mathrm{dof}=398$. The fit parameters, and their respective one standard deviation errors, are determined with very good precision as
\begin{align}
\eta        & = 0.98581\pm0.00024, \nonumber \\
\varepsilon & = 0.00519\pm0.00022, \nonumber \\
\Theta_1    & = -0.03219\pm0.00060.
\end{align}

The residuals are randomly distributed across the full range of the independent variable, as evidenced by the runs test. Yielding a p-value of 0.61, it fails to reject the null hypothesis at a significance level of 0.05. Therefore, we attest robustness of the system to the quantum fluctuations that could have had an effect on the circuit outputs during the $\sim 30$ minutes taken by the experiment.

In Fig.~\ref{fig:A-circ_ibmq} we plot experimental results against theory. Comparing the fitted parameters with Eq.~(\ref{eq:1qb}), $\eta$ shows an amplitude drop most probably due to readout errors but in principle also to the open nature of the system's evolution, while the value for $\Theta_1$ represents systematic calibration errors. The asymmetry introduced by the $\varepsilon$ bias is consistent with both energy dissipation by which the qubit decays from state $\ket{1}$ to $\ket{0}$ and different readout error rates for each of these states.

\subsection{Optimal which-path detection \label{sec:exp-wp}}

This experiment requires the computation of two families of circuits. One applies $P_\mathrm{i}^\dagger=H$ to estimate $\langle X \rangle$ on $\mathrm{q_i}$ and obtain ${\cal V}$, whereas the other is used to find ${\cal D}$ by making $P_\mathrm{i}^\dagger=\mathds{1}$. Both types of circuits carry $P_\mathrm{d}^\dagger=O_\alpha^\dagger$ and were executed with $S=8192$ shots on $C=10201$ pairs of $(\phi,\alpha)$ values sampled on a regular $101\times101$ square grid in $\phi,\alpha\in\left[0,2\pi\right]$. The full experiment therefore included $20402$ circuits in $23$ batches and took $\sim 23$ hours to complete. The batches were prepared with both families of circuits being executed at the same values of the circuit variables in alternating order, shot by shot. The $(\phi, \alpha)$ space was sampled uniformly at random.

The results are analyzed separately for each circuit family. For the first one, we obtained $\overline{X}$ and $\sigma_{\overline{X}}$ on the first qubit as in the previous section. The experimental model approximating Eq.~(\ref{eq:avX}) has four parameters, and the fitting procedure converges with $\chi_\nu^2\approx2.19$. The fit parameters, with one standard deviation error, are
\begin{align}
\eta        & = 0.93039\pm0.00017, & \Theta_1 & = -0.03704\pm0.00026, \nonumber \\
\varepsilon & = 0.03700\pm0.00009, & \Theta_2 & = 0.00512\pm0.00037.
\label{eq:X-q1-exp-params}
\end{align}
Once again, we see no large shifts in calibration with $\Theta_{1,2}$ close to zero. This time, however, we observe a slightly larger drop in $\eta$ as well as a larger $\varepsilon$ bias in comparison with the one-qubit circuit. Assuming the model to be correct, causes might include larger measurement error in the physical qubit used to encode $\mathrm{q_i}$, as well as execution errors introduced by the two-qubit gate, not present in the previous circuit.

From the results obtained with the second family of circuits, distinguishability $\overline{\mathcal{D}}$ is estimated using the relative frequencies of successfully determining the path on $\mathrm{q_i}$. We follow Eq.~(\ref{eq:p_success}) and Eq.~(\ref{eq:p_success_empirical}) to obtain
\begin{equation}
\overline{\mathcal{D}}=\frac{\overline{\mathrm{p}}(00)}{\overline{\mathrm{p}}(00)+\overline{\mathrm{p}}(01)}+\frac{\overline{\mathrm{p}}(11)}{\overline{\mathrm{p}}(10)+\overline{\mathrm{p}}(11)}-1,
\label{eq:Dist-exp}
\end{equation}
and the associated standard error
\begin{equation}
\sigma_{\overline{\mathcal{D}}}=\sqrt{\frac{\overline{\mathrm{p}}(00)~\overline{\mathrm{p}}(01)}{S\left(\overline{\mathrm{p}}(00)+\overline{\mathrm{p}}(01)\right)^{3}}+\frac{\overline{\mathrm{p}}(10)~\overline{\mathrm{p}}(11)}{S\left(\overline{\mathrm{p}}(10)+\overline{\mathrm{p}}(11)\right)^{3}}}.
\label{eq:SE-dist}
\end{equation}
We proceed with the two-dimensional fit of all the estimations with a three-parameter model obtained from Eq.~(\ref{eq:D-purestate}). The procedure yields $\chi_\nu^2\approx1.34$, providing
\begin{align}
\eta        & = 0.90216\pm0.00027, & \Theta_2 & = -0.00224\pm0.00029, \nonumber \\
\varepsilon & = 0.00253\pm0.00023.
\end{align}
Thus, we see no relevant shifts in calibration but a noticeable drop in $\eta$ attributable to the same causes as those for the first family of circuits, with the addition of measurement errors on $\mathrm{q_d}$. The cause for the $\varepsilon$ bias is not as straightforward to single out as before.
\begin{figure}
\includegraphics[width=1\linewidth]{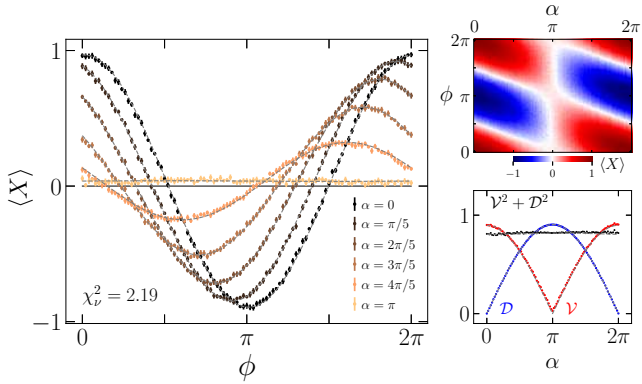}
\caption{Experimental results for the two families of circuits in Fig.\ref{fig:whichpath}b in which the detector qubit is measured in the optimal basis and the inferometer qubit is measured both in the $X$ and the $Z$ basis. Experimental data for $\mathrm{q_i}$ on the left side panel correspond to $\overline{X}\pm2\sigma_{\overline{X}}$. Each of these panels is equivalent to those described in Fig.~\ref{fig:B-circ_insilico}.}
\label{fig:B-circ_ibmq}
\end{figure}

\begin{figure}
\includegraphics[width=1\linewidth]{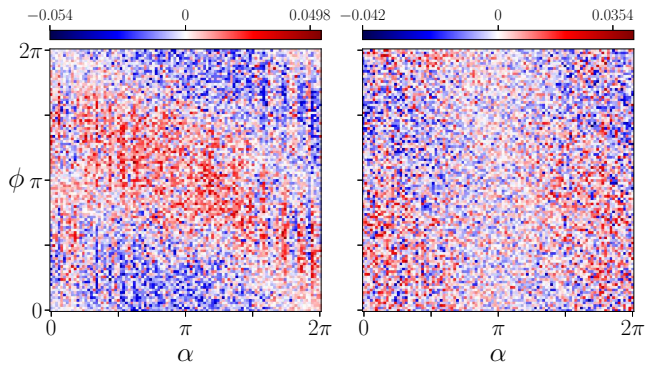}
\caption{Residual error maps $z_j - \hat{z}_j$ for the second experiment. On the left, the residuals of the fit to the $\overline{X}$ model of the first family of circuits are shown. On the right, those of the fit to the $\overline{\cal D}$ model, obtained with the second family of circuits, are shown. The colorbars are marked with the lowest and highest recorded values.}
\label{fig:B-circ_ibmq_res}
\end{figure}

We compute the visibility and distinguishability curves from the two fitted models, obtaining then ${\cal V}^2+{\cal D}^2$. These are represented by the dashed lines in the bottom right panel of Fig.~\ref{fig:B-circ_ibmq} and essentially match the values obtained from the experimental data (in color). However, while the dashed visibility curve obtained from the fitted model reaches ${\cal V} = 0$ by construction, when we look directly at the experimental visibility values (red dots) we find ${\cal V}_\mathrm{min} = 0.030\pm 0.004$ at $\alpha=\pi$ and ${\cal V}_\mathrm{max} = 0.917\pm 0.004$ at $\alpha=1.94\pi$, where the standard deviation error is obtained with $10000$ bootstrap samples. As for the blue experimental data points, for $\overline{{\cal D}}$, they represent the average of the values for $\overline{{\cal D}}$ obtained with all the different circuits for that $\alpha$ value and the different $\phi$.

Comparing this panel with that of Fig.~\ref{fig:B-circ_insilico}, the loss of information is apparent for both the interference visibility ${\cal V}$ and the path distinguishability ${\cal D}$, lowering the amplitudes of Eq.~(\ref{eq:D-purestate}) and Eq.~(\ref{eq:V-purestate}), and modifying Eq.~(\ref{eq:englert-equality}) to ${\cal V}^2+{\cal D}^2 < 1$. This comes about due to the mixing of the final pure state distributions at the output with the noise distribution. In this case, we conclude the noise distribution mostly arises from the errors occurring during the readout process by verifying that the measurement error mitigation protocol closely restores $\eta\approx1$ in both models, and thus ${\cal V}^2+{\cal D}^2 \approx 1$. Since the circuits are so shallow, decoherence does not really play a role in the loss of quantum information here.

The runs tests for the residual errors in the order of execution of the first and the second family of circuits yield p-values numerically close to zero. Therefore, we reject the null hypothesis at the 0.05 significance level and have evidence for autocorrelation in the data. This signals some slight instability in the physical properties of the system, which we examined only to find no significant impact in the presented analysis. Both experimental models are still able to closely fit the data, with residual standard errors, $\mathrm{RSE}\equiv\sqrt{\frac{1}{\mathrm{dof}} \Sigma_{j=1}^C(z_j-\hat{z}_j)^2 }$, of 0.01274 in the first case and 0.00946 in the second case. Scores $R^2\approx0.99926$ and $0.99888$ are obtained for each model, respectively.

More importantly, we find a small but systematic deviation from the theory, mostly in the results from the first family of circuits, but also on those from the second. Regarding the former, the anomaly is already suggested by the fact that $\chi_\nu^2\approx2.19$, but we confirm it by visual inspection of the plot of the fit residuals as a function of $\phi$ and $\alpha$. While an adequate fit would result in a random distribution of residual errors in this space, the left panel of Fig.~\ref{fig:B-circ_ibmq_res} clearly shows a pattern. As for the latter model, the pattern in the right panel of Fig.~\ref{fig:B-circ_ibmq_res} is less visible, in agreement with the lower $\chi_\nu^2\approx1.34$.

Even if of small magnitude, it becomes clear with the resolution of the data that a functional dependence in $\phi$ and $\alpha$ is missing in the experimental models to be able to fully explain the data. In section \ref{sec:RCM} we re-analyze these results in light of an improved theoretical model of the entangling operation.

\subsection{Quantum Eraser \label{sec:exp-eraser}}

Let us now discuss the third experiment in which we ran the circuit in Fig.~\ref{fig:whichpath}b with $P_\mathrm{i}^\dagger=H$ and $P_\mathrm{d}^\dagger=\mathds{1}$. As in the previous section, we choose $S=8192$ and $C=10201$ pairs of $(\phi,\alpha)$ values sampled on a regular $101\times101$ square grid in $\phi,\alpha\in\left[0,2\pi\right]$. The experiment's $10201$ circuits in $12$ batches took $\sim 12$ hours to complete.

As discussed above, our goal is to demonstrate how measuring $\mathrm{q_d}$ in a projection that cannot access the gathered information renders which-path detection useless while recovering full wavelike behavior on $\mathrm{q_i}$. To do so, we separate the readouts on the latter qubit as conditioned by the output $y\in\left\{ 0,1\right\}$ on the detector, to estimate $\langle X_y \rangle$ in Eq.~(\ref{eq:X0conditioned}) and Eq.~(\ref{eq:X1conditioned}) from
\begin{equation}
\overline{X_y} = \frac{\overline{\mathrm{p}}(0y)-\overline{\mathrm{p}}(1y)}{\overline{\mathrm{p}}(0y)+\overline{\mathrm{p}}(1y)}.
\end{equation}
The associated standard error is obtained with
\begin{equation}
\sigma_{\overline{X_y}} = \sqrt{\frac{4~\overline{\mathrm{p}}(0y)~\overline{\mathrm{p}}(1y)}{S\left(\overline{\mathrm{p}}(0y)+\overline{\mathrm{p}}(1y)\right)^3}}.
\end{equation}

\begin{figure}
\includegraphics[width=1\linewidth]{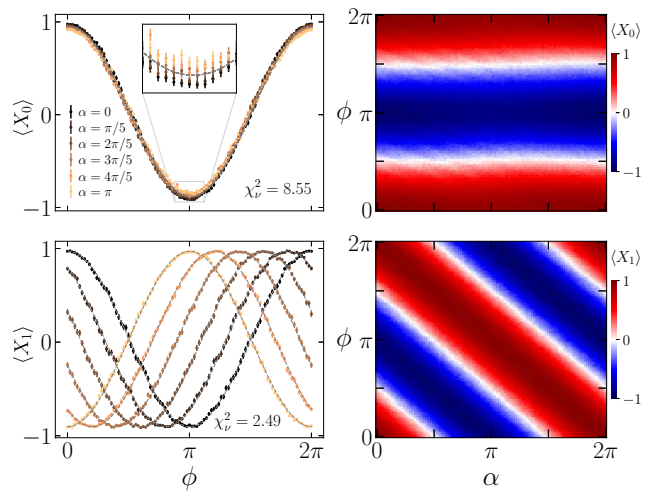}
\caption{Experimental results for the quantum eraser circuit. The expectation values of the $X$ operator on the first qubit broken down by the outcome of the detector are shown, with $\overline{X_0}\pm2\sigma_{\overline{X}_0}$ ($\overline{X_1}\pm2\sigma_{\overline{X}_1}$) in the top (bottom) row. Right-side panels plot the full set of results for the tested $(\phi, \alpha)$ values, while the left-side panels show a selection for six different $\alpha$.}
\label{fig:C-circ_ibmq}
\end{figure}

The results are shown in Fig.~\ref{fig:C-circ_ibmq}, for $\overline{X_0}$ in the top row, and $\overline{X_1}$ in the bottom row. Qualitatively, the full-visibility interference patterns expected in theory are recognizable for both $\overline{X_0}$ and $\overline{X_1}$. To make the analysis quantitative, we first fitted $\overline{X_0}$ data to the expected experimental model obtained from Eq.~(\ref{eq:X0conditioned}). We find
\begin{align}
\eta        & = 0.91740\pm0.00011, & \Theta_1 & = -0.03118\pm0.00020, \nonumber \\
\varepsilon & = 0.03361\pm0.00009,
\end{align}
with $\chi_\nu^2\approx8.55$ and $\mathrm{RSE}=0.03083$. The other subset of the data, for $\overline{X_1}$, was targeted by an experimental model based on Eq.~(\ref{eq:X1conditioned}), achieving $\chi_\nu^2\approx2.49$, $\mathrm{RSE}=0.01885$, and parameters
\begin{align}
\eta        & = 0.93765\pm0.00011, & \Theta_1 & = -0.00565\pm0.00010, \nonumber \\
\varepsilon & = 0.03379\pm0.00009, & \Theta_2 & = -0.00565\pm0.00010.
\end{align}

As can be seen, the values for the fit parameters are very precise in both cases. Furthermore, the runs tests, computed on the chronologically ordered sequence of residuals for both fits separately, fail to reject the null hypothesis for randomness at a significance level of 0.05. Thus, during the execution of this two-qubit circuit, the device had very stable performance.

However, $\chi_\nu^2$ scores show larger deviations from 1, indicating some cause for concern. In fact, the plot of residuals as a function of $(\phi, \alpha)$ shows once again a pattern for both the fits with $\overline{X_0}$ and $\overline{X_1}$ data. The slight disagreement with theory is especially marked for the $\overline{X_0}$ results and can even be seen in the inset plot on the top-left panel of Fig.~\ref{fig:C-circ_ibmq}. A weak dependence in $\alpha$ is apparent, and the visibility of the interference slightly decreases as $\alpha$ approaches $\pi$. This suggests that an imperfection in the implementation of the controlled-$R_\alpha$ gate leaks a small amount of information to the detector qubit which the readout in the $Z$ basis does not destroy, hence slightly reducing visibility on the top qubit. We postpone the analysis of this problem to section \ref{sec:RCM}.

Finally, to understand how effective the detector qubit is at discriminating the paths over the interferometer, we measured distinguishability in this readout basis. To that end, we carried out yet another independent experiment. This fourth one translates $P_\mathrm{i}^\dagger=P_\mathrm{d}^\dagger=\mathds{1}$ in the language of Fig.~\ref{fig:whichpath}b to $G_\mathrm{i}=\mathds{1}$ and $G_\mathrm{d}=U_1(\frac{\alpha}{2})$ in Fig.~\ref{fig:transp-circs}. The same grid with $C=10201$ pairs of $(\phi, \alpha)$ values is probed, as well as the same number of shots $S=8192$.

Distinguishability is again measured from Eq.~(\ref{eq:Dist-exp}) and Eq.~(\ref{eq:SE-dist}). Since $\mathrm{p}_{\mathrm{succ}}$, as computed with Eq.~(\ref{eq:p_success_empirical}), is expected to be 0.5 with the detector readout directly on the states in Eq.~(\ref{eq:detector-states}), theory predicts the \emph{measured} distinguishability \cite{bjork1998complementarity} to be
\begin{equation}
{\cal D}_m=0,
\label{eq:dist-ZZ-circ}
\end{equation}
independently of $\phi$ and $\alpha$. Note that for the previous case of the detector measurement in the optimal basis, the measured distinguishability, as obtained with Eq.~(\ref{eq:p_success}) and Eq.~(\ref{eq:p_success_empirical}), matches the actual distinguishability ${\cal D}$ in Eq.~(\ref{eq:D-definition}) precisely because of the optimal measurement.

Given ${\cal D}_m=0$, we used a simple model to fit the results in which only the $\varepsilon$ parameter from Eq.~(\ref{eq:experimental-model}) is present. The procedure returns $\varepsilon=-0.00370\pm 0.00011$ with $\chi_\nu^2\approx 2.06$ and $\mathrm{RSE}=0.01581$. Inspection of the residuals as a function of $\phi$ and $\alpha$ once again reveals a smooth pattern. Therefore, measured distinguishability is neither zero nor constant across the $(\phi,\alpha)$ space, confirming the conclusion from the previous experiment of a failure to completely erase the quantum information and produce entirely wavelike behavior. This experiment also had no significant environmental perturbations during execution, as evidenced by the p-value of 0.96 for the runs test on the ordered sequence of residuals, failing to reject the null at a significance level of 0.05.

\section{Reassessing the circuit models \label{sec:RCM}}

As we have seen in the previous two-qubit experiments, results were frequently slightly off the theoretical predictions put forward in section \ref{sec:wp-theo}. These anomalies could not be removed with measurement error mitigation, indicating they do not stem from the readout process. We now take into account the systematic deviations between the ideal gates in the presented circuits and the actual hardware-level operations to rederive the theoretical models. We analyze how this allows for a better fit of the experimental results and thus accounts for the moderate, but systematic, deviations from the theory.

\subsection{Accounting for coherent errors}

Take again the operations in the physical circuit of Fig.~\ref{fig:transp-circs}. With the exception of the gates parametrized by $\phi$ and $\alpha$, all other single-qubit gates are applied with fixed input values. Depending on the choices for $G_{\mathrm{i}}$ and $G_{\mathrm{d}}$, any of the circuits will only have $U_{2}$ gates on the top qubit, and both $U_{1}$ and $U_{2}$ gates on the bottom qubit. Let us first model single-qubit gate errors (SQGEs) by adding a bias parameter to each rotation angle of the $U_2$ gate on the interferometer qubit,
\begin{equation}
U_{2}^{(\mathrm{q_i})}\left(x,y\right)\equiv U_{2}\left(x+\theta_{1},y+\theta_{2}\right),
\label{eq:U2-q1}
\end{equation}
and other independent bias parameters to both types of gates on the detector qubit,
\begin{align}
U_{2}^{(\mathrm{q_d})}\left(x,y\right) & \equiv U_{2}\left(x+\theta_{3},y+\theta_{4}\right), \label{eq:U2-q2} \\
U_{1}^{(\mathrm{q_d})}\left(y\right)   & \equiv U_{1}\left(y+\theta_{5}\right). \label{eq:U1-q2}
\end{align}

Considering these three types of gates, with the five additional parameters, to be the actual operations in Fig.~\ref{fig:transp-circs}, we can recalculate output expressions for all the circuits presented before. Namely, Eq.~(\ref{eq:D-purestate}), as computed from Eqs.~(\ref{eq:p_success}) and (\ref{eq:p_success_empirical}), and Eqs.~(\ref{eq:avX}), (\ref{eq:X0conditioned}), (\ref{eq:X1conditioned}), and (\ref{eq:dist-ZZ-circ}). We obtain, respectively,
\begin{equation}
{\cal D} = \sqrt{\cos^{2}\left(\theta_{3}+\theta_{4}\right)\sin^{2}\left(\frac{\alpha}{2}-\theta_{5}\right)},
\label{eq:D-purestate_SQGE}
\end{equation}
\begin{align}
\left\langle X\right\rangle & = \frac{1}{2}\left(\cos\left(\alpha+\phi+\theta_1+\theta_2-\theta_5\right)\right. \nonumber \\
 & \quad \left.+ \cos\left(\phi+\theta_1+\theta_2+\theta_5\right)\right),
\label{eq:avX_SQGE}
\end{align}
\begin{equation}
\left\langle X_{0}\right\rangle = \cos\left(\phi+\theta_1+\theta_2+\theta_5\right),
\label{eq:X0conditioned_SQGE}
\end{equation}
\begin{equation}
\left\langle X_{1}\right\rangle = \cos\left(\alpha+\phi+\theta_1+\theta_2-\theta_5\right),
\label{eq:X1conditioned_SQGE}
\end{equation}
\begin{equation}
{\cal D}_m = 0.
\label{eq:Dm-purestate_SQGE}
\end{equation}

These models provide a microscopic justification for the $\Theta_{1,2}$ parameters introduced with Eq.~(\ref{eq:bigthetas}). As can be seen, $\theta_3$ and $\theta_4$ errors might reduce the maximum attainable ${\cal D}$, while $\theta_1$, $\theta_2$ and $\theta_5$ can only shift the interference pattern for $\left\langle X\right\rangle$ without compromising ${\cal V}$. However, none of these SQGEs explains the slight dependence in $\alpha$ of the interference contrast for $\left\langle X_{0}\right\rangle$, nor the varying ${\cal D}_m$ for different $\phi$ and $\alpha$, both of which have been observed.

To understand what might be missing, let us note that state of the art hardware achieves error rates reaching $10^{-4}$ on single-qubit gates \cite{ibmq}. The daunting challenge, though is to apply the \textsc{cnot} gate with the same level of accuracy. The current implementation of the \textsc{cnot} gate in IBM~Q devices uses the cross-resonance (CR) gate together with single-qubit rotations to reach errors between $10^{-2}$ and $10^{-3}$. Significant work is being devoted to improve the fidelity of the CR gate with carefully designed pulse sequences. These would ideally generate only a $ZX$ interaction term. An echoed CR sequence with target rotary pulses has recently shown improved performance in reducing the undesired error terms in the final pulse Hamiltonian \cite{sundaresan2020reducing}. The \textsc{cnot} used in the experiments we report is based on this sequence.

To model the unitary errors in the \textsc{cnot} implementation derived from the aforementioned CR sequence, we introduce \emph{biased} \textsc{cnot} gate models written up in terms of different adimensional \emph{bias ratios}, $\beta_j$, between the coupling strength of the extra error terms and the desired $ZX$ interaction term. We take into account only the two-qubit subspace of the effective CR pulse Hamiltonian, and the error terms comprising entanglement with spectator qubits are not considered. The simplest introduced gate model, $\textsc{bcnot}_2$, contains only two bias ratios, $\beta_1$ and $\beta_2$, connected to $IY$ and $IZ$ errors. If classical crosstalk from the CR drive accidentally acting on the target qubit is included, error terms for $IX$, $ZY$ and $ZZ$ are also considered, extending the previous model with $\beta_3$, $\beta_4$ and $\beta_5$ to obtain $\textsc{bcnot}_5$:
\begin{widetext}
\begin{align}\left\langle c't'\left|\;\textsc{bcnot}_{5}\;\right|c\,t\right\rangle = & \frac{i\left(c'-1\right)-c c'\left(1+i\right)+ic}{\sqrt{2}}\left(\cos\left(\gamma_{c}\pi/4\right)\left(t\left(1+i\right)\left(2t'-1\right)-\left(1+i\right)t'+i\right)\right. \nonumber\\
 & \left.+\frac{\sin\left(\gamma_{c}\pi/4\right)}{\gamma_{c}}\left(\beta_{1}+\beta_{2}+i\beta_{3}+\left(-1\right)^{c}\left(\beta_{4}+\beta_{5}+i\right)+t\left(1+i\right)\left(-\beta_{1}\right.\right.\right. \nonumber\\
 & \left.\left.\left.+i\left(\beta_{2}+\beta_{3}\left(2t'-1\right)+\left(-1\right)^{c}\left(i\beta_{4}+\beta_{5}+2t'-1\right)\right)\right)\right.\right. \nonumber\\
 & \left.\left.+t'\left(i-1\right)\left(\beta_{1}+i\beta_{2}-\beta_{3}+\left(-1\right)^{c}\left(\beta_{4}+i\beta_{5}-1\right)\right)\right)\right),
\label{eq:BCNOT5}
\end{align}
with a total of five bias parameters, where $c,t,c',t'\in\left\{ 0,1\right\}$ and
\begin{equation}
\gamma_{c}=\sqrt{\left(\beta_{1}+\left(-1\right)^{c}\beta_{4}\right)^{2}+\left(\beta_{2}+\left(-1\right)^{c}\beta_{5}\right)^{2}+\left(\beta_{3}+\left(-1\right)^{c}\right)^{2}},
\label{eq:gamma}
\end{equation}
\end{widetext}
so that Eq.~(\ref{eq:BCNOT5}) encodes an $\mathrm{U}(4)$ operator. In the no bias limit $\beta_{1,\ldots,5}\shortrightarrow 0$, the \textsc{bcnot} gates converge to the ideal \textsc{cnot}, a perfect entangler \cite{zhang2003geometric}. When bias is present, the ability to maximally entangle some acted upon product state is severely hampered or entirely absent. The details of this derivation are provided in Appendix \ref{sec:BCNOT-deriv}.

Our goal now would be to obtain further generalizations of Eqs.~(\ref{eq:D-purestate_SQGE}\,--\,\ref{eq:Dm-purestate_SQGE}) by replacing the ideal \textsc{cnot} in the circuit of Fig.~\ref{fig:transp-circs} first with the $\textsc{bcnot}_2$ alternative and then with $\textsc{bcnot}_5$, in addition to the single-qubit gates in Eqs.~(\ref{eq:U2-q1}\,--\,\ref{eq:U1-q2}). We would then use those ten analytical models, supplemented by $\eta$ and $\varepsilon$ as in Eq.~(\ref{eq:experimental-model}), to fit the experimental data, drawing conclusions about the actual physics taking place in the device.

However, such a program results in very long and cumbersome expressions which are often difficult to compute and simplify to an interpretable form. We could alleviate the problem by alternatively using a truncated series expansion of the $\textsc{bcnot}$ gates in their $\beta_{1,\ldots,5}\approx0$ parameters. Nevertheless, to make sure higher-order contributions are not missed when modeling the experimental results, we consider the exact \textsc{bcnot} gates and fit the data to the readout probabilities obtained by numerically computing the circuits instead.

\subsection{Model evaluation}

Let us now fit the experimental results for ${\cal D}$, $\left\langle X\right\rangle$, $\left\langle X_{0}\right\rangle$, $\left\langle X_{1}\right\rangle$, and ${\cal D}_m$ to the theoretical predictions obtained by numerically calculating the output probabilities of the respective circuits with the different models for single- and two-qubit gates. The obtained $g(\phi,\alpha)$ models are again allowed to scale and shift with $\eta$ and $\epsilon$, following Eq.~(\ref{eq:experimental-model}). We aim at identifying the best model to explain the data from each experiment, and thus understand the origin of the systematic deviations encountered in section \ref{sec:exp}.

For that, we evaluate the $\chi_\nu^2$ and $\mathrm{RSE}$ scores of each fit in order to compare them. However, we do not fit and test a model on the same data, as this could erroneously signal better performance of more complex and parametrized models due to overfitting. We prevent it by evaluating performance on out-of-sample data, using part of the experimental results to fit the model and a different subset to test it. Specifically, we employ tenfold cross-validation with data folds selected uniformly at random in the $(\phi, \alpha)$ space
and score all models on equal splits. This method allows us to identify the models that maximize accuracy and minimize the structure of the residuals while preventing overfitting. We can confidently do so by comparing the assessed measures, reported in Table~\ref{tab:cross-val-metrics}, to rank model adequacy.

Looking at this table, we begin by noting the overall agreement between both scores: the models with the ten single- and two-qubit gate biases, in the last column, more accurately fit the experimental results. The magnitude of improvement is not the same for all of them, though. This is understood on the basis that the measured observables can be more or less resilient to the various coherent errors depending on the circuit, similar to what happens in Eqs.~(\ref{eq:D-purestate_SQGE}\,--\,\ref{eq:Dm-purestate_SQGE}). At times, considering either $\textsc{bcnot}_2$ or $\textsc{bcnot}_5$ gates yields similar performance. However, as seen in the full table of parameters in Appendix~\ref{sec:fit-details}, using $\textsc{bcnot}_5$ yields more stable and reasonable parameter values. Let us take a closer look at each row in Table~\ref{tab:cross-val-metrics} to discuss the details.

First, consider the scores for the $\left\langle X\right\rangle$ models. Table~\ref{tab:cross-val-metrics} clearly shows that taking into account SQGEs and both the $\textsc{bcnot}_2$ and $\textsc{bcnot}_5$ gates maximizes performance. We also find that the obvious pattern in the left panel of Fig.~\ref{fig:B-circ_ibmq_res} mostly disappears to the human eye already using $\textsc{bcnot}_2$ by plotting the residuals from any of the ten splits of the cross-validation procedure. However, simply on the basis of the goodness of fit measures, there is no clear choice between $\textsc{bcnot}_2$ and $\textsc{bcnot}_5$, indicating that there may be some overparametrization of the model obtained with the latter. The ambiguity is reduced by Table~\ref{tab:all-fits-params}, in which less problematic and more realistic fit parameters support the adoption of $\textsc{bcnot}_5$ as the most reliable model for the origin of the data.

With respect to the $\left\langle X_0\right\rangle$ and $\left\langle X_1\right\rangle$ conditional interference patterns, Table~\ref{tab:cross-val-metrics} unambiguously validates SQGEs and the $\textsc{bcnot}_5$ gate as the best performing choice. The goodness of fit improvement is especially significant for $\left\langle X_{0}\right\rangle$. In fact, plotting again the region represented in the inset of the top-left panel of Fig.~\ref{fig:C-circ_ibmq} with the fitted equations, we observe that the single ideal sinusoidal curve is replaced by multiple ones adjusting the points from each value of $\alpha$. Even so, in the residuals of both $\left\langle X_{0}\right\rangle$ and $\left\langle X_{1}\right\rangle$ fits, patterns are still visually clear, albeit of diminished magnitude. In particular, these patterns show high frequency features with striking clarity, which would require further understanding. If we look closely at the bottom-left panel of Fig.~\ref{fig:C-circ_ibmq} for instance, we can actually see these very small oscillations of the data around the sinusoidal curves of that fit.

This observation that some small additional effect is not being taken into account by the models leads us well into the discussion of ${\cal D}$ and ${\cal D}_m$. As can be seen, there is little difference in using any of the models to fit the experimental results of both these quantities, although considering SQGEs and both $\textsc{bcnot}_2$ and $\textsc{bcnot}_5$ slightly improves goodness of fit for the case of ${\cal D}_m$. It is important to note however, that little further improvement is possible to the scores of the simplest ${\cal D}$ model, since these are already quite low to begin with. In fact, these scores are the best ones in the table. Nevertheless, we are also able to identify an odd feature in the plots of the experimental data for both ${\cal D}$ and ${\cal D}_m$: a smooth dependence on $\phi$. This is unexpected because there is no $\phi$ dependence in the long and exact closed-form expressions of the models for these two quantities which integrate SQGEs and the $\textsc{bcnot}_5$ gate. Once again, the data contain evidence of a small additional effect that is not included in our more complex models.

\begin{table}%[t]
\begingroup\setlength{\fboxsep}{0pt}
\colorbox{gray!5}{\parbox{1\columnwidth}{
\begin{tabularx}{\columnwidth}{ *{6}{m{0.09\columnwidth}>{\centering}m{0.09\columnwidth}>{\centering}m{0.18\columnwidth}>{\centering}m{0.18\columnwidth}>{\centering}m{0.18\columnwidth}>{\centering\arraybackslash}m{0.18\columnwidth}} }
\hline
\hline
\noalign{\smallskip}
 & & \multicolumn{1}{c}{No SQGEs} & \multicolumn{3}{c}{SQGEs} \\ \noalign{\smallskip} \cline{4-6}
\noalign{\smallskip}
 & & $\textsc{cnot}$ & $\textsc{cnot}$ & $\textsc{bcnot}_{2}$ & $\textsc{bcnot}_{5}$ \\
\hline
\hline
\noalign{\smallskip}
${\cal D}$ & $\begin{array}{l} \chi_\nu^2\\ \mathrm{RSE}\end{array}$ & $\begin{array}{r} \;\;\;1.34692 \\ 0.00950 \end{array}$ & $\begin{array}{r} \;\;\;1.34766 \\ 0.00950 \end{array}$ & $\begin{array}{r} 1.34993 \\ 0.00950 \end{array}$ & $\begin{array}{r} 1.35376 \\ 0.00952 \end{array}$ \\
\hline
\noalign{\smallskip}
$\left\langle X\right\rangle$ & $\begin{array}{l} \chi_\nu^2\\ \mathrm{RSE}\end{array}$ & $\begin{array}{r} \;\;\;5.78430 \\ 0.02062 \end{array}$ & $\begin{array}{r} \;\;\;2.20479 \\ 0.01278 \end{array}$ & $\begin{array}{r} \mathbf{1.81458} \\ \mathbf{0.01161} \end{array}$ & $\begin{array}{r} \mathbf{1.81619} \\ \mathbf{0.01161} \end{array}$ \\
\hline
\noalign{\smallskip}
$\left\langle X_0\right\rangle$ & $\begin{array}{l} \chi_\nu^2\\ \mathrm{RSE}\end{array}$ & $\begin{array}{r} \;11.07644 \\ 0.03670 \end{array}$ & $\begin{array}{r} \;\;\;8.61654 \\ 0.03094 \end{array}$ & $\begin{array}{r} 7.50310 \\ 0.02775 \end{array}$ & $\begin{array}{r} \mathbf{2.71485} \\ \mathbf{0.02016} \end{array}$ \\
\hline
\noalign{\smallskip}
$\left\langle X_1\right\rangle$ & $\begin{array}{l} \chi_\nu^2\\ \mathrm{RSE}\end{array}$ & $\begin{array}{r} \;\;\;2.83404 \\ 0.01997 \end{array}$ & $\begin{array}{r} \;\;\;2.51185 \\ 0.01891 \end{array}$ & $\begin{array}{r} 2.48586 \\ 0.01873 \end{array}$ & $\begin{array}{r} \mathbf{2.42447} \\ \mathbf{0.01865} \end{array}$ \\
\hline
\noalign{\smallskip}
${\cal D}_{m}$ & $\begin{array}{l} \chi_\nu^2\\ \mathrm{RSE}\end{array}$ & $\begin{array}{r} 2.05986 \\ 0.01582 \end{array}$ & $\begin{array}{r} 2.07046 \\ 0.01586 \end{array}$ & $\begin{array}{r} \mathbf{2.02330} \\ \mathbf{0.01568} \end{array}$ & $\begin{array}{r} \mathbf{2.02358} \\ \mathbf{0.01568} \end{array}$ \\
\hline
\hline
\end{tabularx}
}}\endgroup
\caption{Average performance measures for the fits obtained with different single- and two-qubit gate models, as evaluated with tenfold cross-validation. Boldface highlights the best-performing models. Table~\ref{tab:all-fits-params} presents the parameter values.}
\label{tab:cross-val-metrics}
\end{table}

Therefore, while definitely improving goodness of fit as compared to the results in section \ref{sec:wp-theo}, these new and more parametrized models still lack some of the physics that produces the results. This does not change if we consider a more complete model for SQGEs in which, on top of the biases introduced in Eqs.~(\ref{eq:U2-q1}\,--\,\ref{eq:U1-q2}), fit parameters factored with $\phi$ and $\alpha$ are also added. This would be an attempt to capture an eventual linear dependence of the errors on those variables appearing in the nonconstant gates of Fig.~\ref{fig:transp-circs}. However, it provides no significant improvement to the performance of the models in the last three columns of Table~\ref{tab:cross-val-metrics}.

As such, we conjecture two possibilities for additional effects that could improve the leftover lack of fit and are not taken into account by the $\textsc{bcnot}$ gates based on the CR gate analysis in \cite{sundaresan2020reducing}. These are the potentially significant \emph{spectator errors}, consisting of couplings to nearest neighbor qubits, and the non-negligible errors due to the crosstalk coming back to the control transmon from the rotary tone of the CR gate.

Regardless of that, cross-validation shows that the more complete description of the physical operations introduced above greatly improves the modeling of the data, capturing therefore most of the physics in an interpretable manner. Moreover, the experimental observation of the inability to achieve the expected situations of full particlelike and full wavelike statistics can be explained as an important consequence of the biased \textsc{cnot} operation, as shown promptly below.

\subsection{Limitations on extreme-case statistics imposed by the biased CNOT \label{sec:BCNOT-limitations}}

As we can see from Eqs.~(\ref{eq:D-purestate_SQGE}\,--\,\ref{eq:Dm-purestate_SQGE}), SQGEs cannot prevent $\left\langle X\right\rangle$ from achieving ${\cal V}=0$ and $1$, since Eq.~(\ref{eq:V-purestate}) would only suffer a phase shift. Similarly, SQGEs alone cannot compromise ${\cal V}=1$ in $\left\langle X_0\right\rangle$ and $\left\langle X_1\right\rangle$ interference patterns, which would also only experience a phase shift, leaving contrast unchanged. Moreover, ${\cal D}_m$ is also unaffected by $\theta_{1,\ldots,5}$ errors. The only exception is ${\cal D}$, which can actually fall short of achieving ${\cal D}=1$ because of SQGEs. However, in this last case, we cannot tell from the experiment whether this is due to SQGEs or incoherent noise accounted for by $\eta$.

Here, we quickly show that the biased \textsc{cnot} gate introduced in Eq.~(\ref{eq:BCNOT5}) can have this limiting effect, thus preventing the observation of full particlelike and wavelike statistics even with ideal single-qubit gate execution. To demonstrate that, we again recompute the output probabilities of all the circuits studied before, this time by taking the full expression for $\textsc{bcnot}_5$ as replacement of the \textsc{cnot} in Fig.~\ref{fig:transp-circs} while preserving single-qubit gates as in Eq.~(\ref{eq:ibm-base-ops}). From those, all other quantities follow.

Before attempting an exact solution, it is instructive to examine the series expansion of the models truncated to first order in all $\beta_j\approx0$, with $j=1,\ldots,5$. These five models derived from the four different circuits are:
\begin{widetext}
\begin{equation}
\mathcal{D}=\sin\left(\frac{\alpha}{2}\right)-\left(\beta_1+\beta_2+\beta_4+\beta_5\right)\cos\left(\frac{\alpha}{2}\right),
\label{eq:D-BCNOT5-1}
\end{equation}
\begin{equation}
\left\langle X\right\rangle = \frac{\left(1+\beta_2-\beta_4\right)\cos\phi+\left(1-\beta_2+\beta_4\right)\cos\left(\alpha+\phi\right)}{2} -\frac{\beta_1+\beta_2+\beta_4+\beta_5}{2}\left(\sin\phi-\sin\left(\alpha+\phi\right)\right),
\label{eq:X-BCNOT5-1}
\end{equation}
\begin{equation}
\left\langle X_{0}\right\rangle =\cos\left(\phi\right)+\sin\left(\phi\right)\left(\left(\beta_1-\beta_5\right)\sin\left(\frac{\alpha}{2}\right)-\beta_1-\beta_2-\beta_4-\beta_5\right),
\label{eq:X0-BCNOT5-1}
\end{equation}
\begin{equation}
\left\langle X_{1}\right\rangle =\cos\left(\alpha+\phi\right)+\sin\left(\alpha+\phi\right)\left(\left(\beta_1-\beta_5\right)\sin\left(\frac{\alpha}{2}\right)+\beta_1+\beta_2+\beta_4+\beta_5\right),
\label{eq:X1-BCNOT5-1}
\end{equation}
\begin{equation}
\mathcal{D}_{m}=-\beta_1+\beta_5+\left(\beta_2-\beta_4\right)\cos\left(\frac{\alpha}{2}\right)-\frac{\pi}{2}\beta_3\sin\left(\frac{\alpha}{2}\right).
\label{eq:Dm-BCNOT5-1}
\end{equation}
\end{widetext}

The first thing to note is that there is a lot of overparametrization in these expressions. In fact, the first four models could be fitted by considering only nonzero $\beta_1$ and $\beta_2$ parameters. Moreover, in some of these equations, the limitations on the measurement statistics can already be confirmed. For instance, Eq.~(\ref{eq:X-BCNOT5-1}) cannot exhibit ${\cal V}=0$. However, understanding the extreme values for ${\cal D}$ and ${\cal V}$ is mostly inconclusive using these approximations and requires an exact approach.

\begin{figure}
\includegraphics[width=1\linewidth]{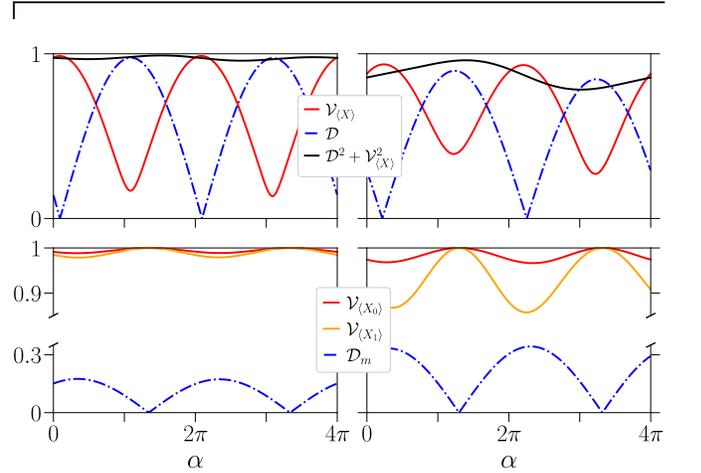}
\caption{Exact visibility and distinguishability plots for the different circuits encoded in Fig.~\ref{fig:transp-circs}, with the \textsc{cnot} gate replaced by a $\textsc{bcnot}_5$ with example bias values. In all of the panels, $\beta_1=0.06$, $\beta_3=-0.05$, $\beta_4=-0.07$, and $\beta_5=0.06$. While $\beta_2=0.09$ on the left-side panels, it is drastically increased to $0.3$ on the right-side ones to emphasize the effect.}
\label{fig:vis_dis_bias}
\end{figure}

As previously stated, calculating complete closed-form models for these circuits results in exceedingly long and cumbersome expressions. Instead, we employ numerics to compute all circuits and derived expressions exactly. The limitations to full particlelike and wavelike statistics imposed by $\textsc{bcnot}_5$ become apparent for all the circuits we tested. The severity of the effect varies with the magnitude of the different bias ratios, but it starts continuously as soon as there is some nonzero $\beta_j$. The only exception occurs for ${\cal V}_{\left\langle X\right\rangle}$, in which $\beta_3$ alone does not deform the function or prevent us from reaching the extremes. We illustrate this with an example in Fig.~\ref{fig:vis_dis_bias}, which plots visibility and distinguishability obtained with the different circuits by specifying some arbitrarily chosen values for the biases.

On the top row, we put together the visibility obtained for $\left\langle X\right\rangle$ with the associated distinguishability, as obtained with the optimal measurement of the detector. As shown, even with only a little bias present, the interferometer qubit cannot display ${\cal V}=0$ or $1$, and the detector measurement cannot achieve ${\cal D}=1$. This effect becomes more perceptible as the bias is increased. Note that ${\cal D}$ is no longer the actual distinguishability but what we would think the distinguishability is by using the $O_{\alpha}^\dagger$ basis measurement. It would be more accurate to name it ${\cal D}_m'$ but we keep ${\cal D}$ to avoid confusion with notation.

On the bottom row of Fig.~\ref{fig:vis_dis_bias}, the visibility of the conditional interference patterns $\left\langle X_0\right\rangle$ and $\left\langle X_1\right\rangle$, which should ideally be $1$, is plotted. In this case, ${\cal V}\approx1$ for two $\alpha$ values within the $4\pi$ period, but all other interference patterns clearly have ${\cal V}<1$. The decrease in visibility as $\alpha$ approaches $\pi$ from zero is consistent with the experimental observations in Fig.~\ref{fig:C-circ_ibmq}. In addition, we also plot ${\cal D}_m$, which confirms via a different route that the readout in the $Z$ basis cannot completely erase two-qubit correlations and produce full wavelike behavior. Note that the $4\pi$ period comes from the range of the input variables to the gates in Fig.~\ref{fig:transp-circs}, which is $2\pi$.

These consequences to extreme-case statistics are thus in alignment with our experimental results. Recall that in the optimal which-path detection experiments we could not confirm ${\cal V}=0$ for any value of $\alpha$, corresponding to full particlelike dynamics, and in the quantum eraser experiments we could not obtain full wavelike statistics since we did not observe constant high ${\cal V}$ or ${\cal D}_m=0$ for all $\alpha$. Therefore, biased two-qubit operations may be a fundamental limitation in realizing Bohr's strong formulation of complementarity, and experimental tests should not overlook eventual small systematic deviations.

\section{Discussion and conclusions \label{sec:disc}}

In this paper, we set out to test interferometric complementarity using a superconducting transmon system. To do so, we proposed a general two-qubit quantum circuit to implement which-path detection with tunable sensitivity and adjustable readout basis. This circuit can be executed in any gate-based quantum computer with a universal gate-set, regardless of the underlying hardware.

We considered four specific choices of the readout basis and thus introduced four different parametrized circuits, which were theoretically analyzed. First, we looked at two circuits in which the detector qubit is measured in the optimal guess basis. This allows capturing the maximum amount of the stored which-path information, optimizing the probability of guessing the correct path through the interferometer independently of its readout basis. We measured the interferometer qubit both in the $X$ and the $Z$ basis. By combining the results from the two circuits, the duality relation was tested for different visibility and distinguishability values, controlled by the detector sensitivity parameter $\alpha$.

The other two circuits we studied measure the detector qubit in the computational basis directly, implementing the so-called quantum eraser setup. In this case, one expects the which-path information to be destroyed at readout of the detector for any value of $\alpha$, obtaining a constant and null measured distinguishability. We have shown that by using the guidance of the detector to sort the measurement results on the interferometer, two full-visibility interference patterns should be recovered, something that is not possible with the optimal readout.

These circuits were then implemented in an IBM~Q quantum system controlled remotely. At a coarse level, we have found good agreement between the experiments and the theory, which is expected given the low depth of the quantum circuits. Interestingly, however, some small but persistent systematic deviations from the theoretical analysis have also been identified at a finer resolution.

Specifically, there was always some degree of waviness in the $X$ basis measurements of the interferometer qubit for all $\alpha$. This has prevented the observation of complete particlelike statistics, contrary to what was expected. Similarly, a complete wavelike behavior of the interferometer was not reached for all $\alpha$ with the quantum eraser circuits. We have found the visibility of the conditional interference patterns to depend on $\alpha$, as well as a residual amount of measured distinguishability across the $\phi$ and $\alpha$ space, both characters of particlelike dynamics. This means that we could devise a way to win the which-path guessing game with a probability greater than 50\% even using the quantum eraser setup, provided we can experiment with the circuit to tune our strategy before we bet.

We were able to relate these deviations between theory and experiments to coherent errors in single-qubit and \textsc{cnot} gates. By properly modeling both types of operations in the physical circuits, the latter of which is based on the cross-resonance gate implementation, we have upgraded the theoretical models and showed these improve data explainability. In doing so, we have demonstrated that it is very difficult to actually generate full wavelike and particlelike statistics, since it requires bias-free two-qubit gates in our circuits.

This understanding raises the question of whether this limitation is exclusive to our implementation and how can nonbiased perfect two-qubit entanglers be operated to overcome it, a pertinent observation given how difficult it appears to be to precisely obtain maximal entanglement. Confidently realizing this prediction of quantum theory would allow us to address Bohr's strong principle of complementarity and not only the weaker duality relation in Eq.~(\ref{eq:VisDis}). Of course, in order to resolve the issue, one must be able to test the circuits with a level of precision in the measured observables and resolution in the variable space that can pierce through very small systematic deviations.

One option to correct the bias without attempting to improve gate calibration would be to customize circuits \cite{bremmer2002practical, zhang2003exact, dawson2006solovaykitaev}. However, adding circuit depth makes the system more susceptible to incoherent noise, leading to results farther from the pure state limit. Moreover, our purposes were not to estimate observables by mixing measurements obtained with different circuits or qubit pairs \cite{wallman2016noise}, which also trades coherent for incoherent errors. Therefore, our findings also show that which-path experiments can be a way to benchmark quantum hardware.

The which-path experiments discussed here were initially proposed as {\em gedanken experiments} to highlight counter-intuitive concepts in quantum theory. Their experimental verification required direct access to sophisticated equipment, available only to a few groups worldwide. With the advent of cloud-based quantum computers, our paper provides a way to remotely test this fundamental principle of quantum mechanics in the circuit model language of quantum computing. The simplicity of the procedure makes it even suitable for undergraduate quantum mechanics courses.

\begin{acknowledgments}
P. M. Q. C. acknowledges Funda\c{c}\~{a}o para a Ci\^{e}ncia e a Tecnologia (FCT) Grant No. SFRH/BD/150708/2020. J. F.-R. acknowledges financial support from FCT Grant No. UTAP-EXPL/NTec/0046/2017, Generalitat Valenciana Grant No. Prometeo2017/139 and MINECO Grant No. PID2019-109539GB. We acknowledge interesting discussions with E. Galv\~ao. We thank IBM for the opportunity to perform this experiment.
\end{acknowledgments}

\appendix

\section{The biased-CNOT gate model \label{sec:BCNOT-deriv}}

In the following, we derive the biased-\textsc{cnot} gate used to model the physical \textsc{cnot} in the device. Let us start by considering a $ZX$ interaction Hamiltonian, with propagator \mbox{$U_{\mathrm{ZX}}\left(\theta\right) = \exp\left(-i\frac{\theta}{2}ZX\right)$}. For $\theta=\pi/2$, it yields
\begin{equation}
U_{\mathrm{ZX}}\left(\frac{\pi}{2}\right) = \frac{1}{\sqrt{2}} \left[\begin{array}{cccc}
1 & -i & 0 & 0\\
-i & 1 & 0 & 0\\
0 & 0 & 1 & i\\
0 & 0 & i & 1
\end{array}\right]\equiv U_{\mathrm{ZX;\frac{\pi}{2}}},
\label{eq:pentangler}
\end{equation}
which is locally equivalent to a \textsc{cnot} and capable of maximally entangling the product state \mbox{$(\ket{00}+\ket{10})/\sqrt{2}$}. Even if we do not know the exact single-qubit rotations that together with $U_{\mathrm{ZX;\frac{\pi}{2}}}$ give the \textsc{cnot}, we can capture them by solving $\textsc{cnot}=U_{\mathrm{ZX;\frac{\pi}{2}}}\cdot M$ for $M$ to obtain
\begin{equation}
M = \frac{1}{\sqrt{2}} \left[\begin{array}{cccc}
1 & i & 0 & 0\\
i & 1 & 0 & 0\\
0 & 0 & -i & 1\\
0 & 0 & 1 & -i
\end{array}\right],
\end{equation}
obeying $\left[M,U_{\mathrm{ZX;\frac{\pi}{2}}}\right]=0$. One way of decomposing $M$ would be with the $S^{\dagger}$ and $R_{x}(-\pi/2)$ single-qubit gates, applied on the control and target qubits, respectively:
\begin{equation}
M=S^{\dagger}\otimes R_{x}\left(-\frac{\pi}{2}\right).
\label{eq:M_decomp}
\end{equation}

Ideally, the cross-resonance gate can be used to simulate an effective $U_{\mathrm{ZX}}\left(\theta\right)$ coupling. It consists in a microwave drive of the control transmon at the frequency of the target, rotating the state of the latter depending on the state of the former. The CR tone has been studied and modeled by an effective Hamiltonian in the Pauli basis \cite{sheldon2016procedure, malekakhlagh2020first}, and multipulse echo sequences and cancellation tones were designed to isolate as much as possible the $ZX$ term. On IBM Q processors, the echoed CR with a target rotary pulse sequence achieves the current state-of-the-art performance in reducing the undesired error terms in the final interaction Hamiltonian \cite{sundaresan2020reducing}, given by
\begin{equation}
H_{\mathrm{eff}}=\tilde{\nu}_{ZX}\frac{ZX}{2}+H_{\mathrm{err_{1}}}+H_{\mathrm{err}_{2}}.
\label{eq:Heff}
\end{equation}
Here,
\begin{equation}
H_{\mathrm{err_{1}}}=\tilde{\nu}_{IY}\frac{IY}{2}+\tilde{\nu}_{IZ}\frac{IZ}{2}
\label{eq:Herr1}
\end{equation}
contains the error terms produced by the standard CR two-pulse echo sequence in the two-qubit subspace, and
\begin{equation}
H_{\mathrm{err_{2}}}=\tilde{\nu}_{IX}\frac{IX}{2}+\tilde{\nu}_{ZY}\frac{ZY}{2}+\tilde{\nu}_{ZZ}\frac{ZZ}{2}
\label{eq:Herr2}
\end{equation}
contains those arising from classical crosstalk from the CR drive accidentally acting on the target qubit. The error terms comprising entanglement with spectator qubits are not being considered.

If $H_{\mathrm{err_{1}}}$ and $H_{\mathrm{err_{2}}}$ could be made vanishingly small, we would implement the perfect entangler in Eq.~(\ref{eq:pentangler}) with $\tau=\frac{\pi}{2\tilde{\nu}_{ZX}}$ in $\exp\left(-i\tau H_{\mathrm{eff}}\right)$. However, when considering the full Hamiltonian, a systematic deviation arises. We can write the resulting block-diagonal operator, $U_{\mathrm{eff};\frac{\pi}{2}}\left(\mathbf{B}\right)$, in terms of five bias parameters $\beta_{1}\equiv\frac{\tilde{\nu}_{IY}}{\tilde{\nu}_{ZX}}$, $\beta_{2}\equiv\frac{\tilde{\nu}_{IZ}}{\tilde{\nu}_{ZX}}$, $\beta_{3}\equiv\frac{\tilde{\nu}_{IX}}{\tilde{\nu}_{ZX}}$, $\beta_{4}\equiv\frac{\tilde{\nu}_{ZY}}{\tilde{\nu}_{ZX}}$, and $\beta_{5}\equiv\frac{\tilde{\nu}_{ZZ}}{\tilde{\nu}_{ZX}}$, taken together as $\mathbf{B}\equiv \left\{ \beta_{1},\beta_{2},\beta_{3},\beta_{4},\beta_{5}\right\}$. The entries of this $4\times4$ matrix can be compactly encoded by
\begin{widetext}
\begin{align}
\left\langle c't'\left|U_{\mathrm{eff};\frac{\pi}{2}}\left(\mathbf{B}\right)\right|c\,t\right\rangle = & \left(c+c'-1\right)^{2}\left(\cos\left(\gamma_{c}\frac{\pi}{4}\right)\left(t+t'-1\right)^{2}+\frac{\sin\left(\gamma_{c}\pi/4\right)}{\gamma_{c}}\left(i\left(t+t'-1\right)\left(\beta_{2}+\left(-1\right)^{c}\beta_{5}\right)\right.\right.\nonumber \\
 & \left.\left.-\left(t+t'-2tt'\right)\left(i\left(\beta_{3}+\left(-1\right)^{c}\right)+\left(-1\right)^{t'}\left(\beta_{1}+\left(-1\right)^{c}\beta_{4}\right)\right)\right)\right),
\end{align}
\end{widetext}
where $c,t,c',t'\in\left\{ 0,1\right\}$ and $\gamma_c$ is written as in Eq.~(\ref{eq:gamma}). One can check that $\lim_{\mathbf{B}\rightarrow \mathbf{0}}U_{\mathrm{eff};\frac{\pi}{2}}\left(\mathbf{B}\right)=U_{\mathrm{ZX;\frac{\pi}{2}}}$.

We now apply the same local operation, $M$, to this result to obtain the physical approximation to the \textsc{cnot}. However, because $\left[M,U_{\mathrm{eff};\frac{\pi}{2}}\left(\mathbf{B}\right)\right]\neq0$, an assumption is required on the ordering of the operators. If we define $\textsc{bcnot}_5 \equiv U_{\mathrm{eff};\frac{\pi}{2}}\left(\mathbf{B}\right)\cdot M$, the result is in Eq.~(\ref{eq:BCNOT5}).

Now suppose we had chosen the definition $\textsc{bcnot}\,'_5 \equiv M\cdot U_{\mathrm{eff};\frac{\pi}{2}}\left(\mathbf{B}\right)$ instead. We would have obtained the same operator we get by mapping $\left\{ \beta_1 \shortrightarrow \beta_2, \; \beta_2 \shortrightarrow -\beta_1, \; \beta_3 \shortrightarrow \beta_3, \; \beta_4 \shortrightarrow  \beta_5, \; \beta_5 \shortrightarrow -\beta_4 \right\} $ in $\textsc{bcnot}_5$. Therefore, while different, the parametrizations of $\textsc{bcnot}\,'_5$ and $\textsc{bcnot}_5$ are connected. This implies that if we use any of these definitions to obtain models to fit experimental data, both options will be able to adjust it and only the interpretation of each fit parameter changes. This suffices for our goal, which is not to fully characterize each bias individually but only to confirm the whole set can model the data.

The same applies to the only two other ways of obtaining the \textsc{cnot} by combining the target entangler with phase and $X$ rotation gates:
\begin{equation}
\small{
\textsc{bcnot}''_5 \equiv \left(S^{\dagger}\otimes \mathds{1}\right)\cdot U_{\mathrm{eff};\frac{\pi}{2}}\left(\mathbf{B}\right)\cdot\left(\mathds{1}\otimes R_{x}\left(-\frac{\pi}{2}\right)\right),
}
\end{equation}
\begin{equation}
\small{
\textsc{bcnot}'''_5 \equiv \left(\mathds{1}\otimes R_{x}\left(-\frac{\pi}{2}\right)\right)\cdot U_{\mathrm{eff};\frac{\pi}{2}}\left(\mathbf{B}\right)\cdot\left(S^{\dagger}\otimes \mathds{1}\right).
}
\end{equation}
For the first option, we have $\textsc{bcnot}''_5 = \textsc{bcnot}_5$, while for the second one we again obtain the same $\textsc{bcnot}_5 \rightarrow \textsc{bcnot}'''_5$ mapping we got for $\textsc{bcnot}'_5$. Therefore, it is also important to note that these maps from $\textsc{bcnot}_5$ to $\textsc{bcnot}'_5$ and $\textsc{bcnot}'''_5$ do not mix $\beta_j$ parameters from the different error Hamiltonians in Eqs.~(\ref{eq:Herr1}, \ref{eq:Herr2}). This is helpful if we want to ignore one of these in the analysis, as it guarantees these four sequences to arrive at a biased \textsc{cnot} gate still connected among each other. Our $\textsc{bcnot}_2$ model takes $H_{err_2}=0$.

\section{Further details on model evaluation \label{sec:fit-details}}

As described in section \ref{sec:RCM}, we take different combinations of ideal and faulty models for the physical gates in Fig.~\ref{fig:transp-circs} to compute four models of the output probabilities of each circuit. We then fit the relevant quantities by allowing parameter optimization within the fixed bounds $\left|\beta_j\right| \leq0.5$, $\left|\theta_k\right| \leq\pi/10$, $\eta \in [0.7,1]$, $\left|\varepsilon\right| \leq0.5$. We employ tenfold cross-validation to assess goodness of fit as well as to find the optimal parameters by averaging them over the different splits used by the algorithm. Precision errors are given by the standard deviation of these ten values.

Consider the optimal parameters in the first column of Table~\ref{tab:all-fits-params} concerning the models using the ideal \textsc{cnot} and no SQGEs. The precision margins in this case are among the best in the table. This is simply because, being less parametrized, these models quickly lock to the data and do not allow much freedom in optimizing the function. However, as seen in Table~\ref{tab:cross-val-metrics}, they yield the worst goodness of fit measures.

In the next three columns of Table~\ref{tab:all-fits-params}, with the more complex models, the optimization returned some final parameter values right at the predefined bounds. Consider the second column. As explained, these fits are performed using numerical calculations as reference, but in this particular case the closed-form expressions are simple enough, being given by Eqs.~(\ref{eq:D-purestate_SQGE}\,--\,\ref{eq:Dm-purestate_SQGE}). Looking at the results in the table, we see that the parameters with unstable values (underlined) are actually not present in those expressions. Thus, it is unsurprising the fit could not optimize them, and these can be ignored.

Now for the third and fourth columns, there are two overlapping reasons contributing to some unstable parameter values. The first is the same as before: some of these error parameters, though possibly existing in the gates in which they were introduced, might not reflect in the observables being measured by different circuits. However, by calculating the circuit numerically we cannot tell which ones, since all parameters are included with each gate to obtain the most general possible output; we do not add the additional layer of complexity to try to numerically understand which gate parameters could be dropped for each particular circuit.

The second reason is the fact that, as described in section \ref{sec:RCM}, the data carry evidence that coherent behavior persists unaccounted for by the more complex models we considered. Since these models have more degrees of freedom to optimize the fit and the data are not entirely compatible with the model, the parameters adapt in ways that might not reflect the physics from which they were derived. Nevertheless, it is clear that the best overall results in terms of goodness of fit as well as less problematic values for the optimal parameters are obtained with the models considering SQGEs and the $\textsc{bcnot}_5$ gate.

Note that, in general, $\beta_j$ and $\theta_k$ parameters shown in different rows of Table~\ref{tab:all-fits-params} need not match because these correspond to different experiments, performed with different calibrations of the device as well as different qubits. However, for the models of $\left\langle X\right\rangle$ and ${\cal D}$, as well as $\left\langle X_0\right\rangle$ and $\left\langle X_1\right\rangle$, for which data were obtained at the same execution time in the device, it seems some of the $\beta_j$ and $\theta_k$ values in the last column of Table~\ref{tab:all-fits-params} are comparable while others are not. This inconsistency is in agreement with what we argued before: there is still some physical effect unaccounted for, which gives some freedom to the different models to find different optimal values for the parameters included in the model. Furthermore, if overparameterization is present, the final models are also optimizable with different values of the parameters, which could also explain some differences.

To summarize, we have shown that our models capture most of the physics of the operations, doing a very good job at modeling the data in an interpretable manner. However, we are also sure there is still some small degree of incompleteness in them, not only because of the remnant lack of fit and nonuniform distribution of residuals, but also due to some unstable fit parameters.

\begin{table}[ht]
\footnotesize
\begin{minipage}[c]{\textwidth}
\begin{tabularx}{\textwidth}{ *{6}{m{0.05\textwidth}>{\centering}m{0.05\textwidth}>{\centering}m{0.213\textwidth}>{\centering}m{0.213\textwidth}>{\centering}m{0.213\textwidth}>{\centering\arraybackslash}m{0.213\textwidth}} }
\hline
\hline
\noalign{\smallskip}
 & & \multicolumn{1}{c}{No SQGEs} & \multicolumn{3}{c}{SQGEs} \\ \noalign{\smallskip} \cline{4-6}
\noalign{\smallskip}
 & & $\textsc{cnot}$ & $\textsc{cnot}$ & $\textsc{bcnot}_{2}$ & $\textsc{bcnot}_{5}$ \\
\hline
\hline
\noalign{\smallskip}
${\cal D}$ & $\begin{array}{l} \eta\\ \varepsilon\\ \theta_1\\ \theta_2\\ \theta_3\\ \theta_4\\ \theta_5\\ \beta_1\\ \beta_2\\ \beta_3\\ \beta_4\\ \beta_5 \end{array}$ & $\begin{array}{ll} \;\;\;0.90206 &\pm 0.00011\\ \;\;\;0.00261 &\pm 0.00009\\ \qquad-\\ \qquad-\\ \qquad-\\ \qquad-\\ \qquad-\\ \qquad-\\ \qquad-\\ \qquad-\\ \qquad-\\ \qquad- \end{array}$ & $\begin{array}{ll} \;\;\;0.91506 &\pm 0.00453\\ \;\;\;0.00261 &\pm 0.00011\\ \;\;\;\underline{0.01839\,\pi} &\pm 0.06088\,\pi\\ -\underline{0.06065\,\pi} &\pm 0.07914\,\pi\\ \;\;\;0.02689\,\pi &\pm 0.00462\,\pi\\ \;\;\;0.02604\,\pi &\pm 0.00430\,\pi\\ \;\;\;0.00037\,\pi &\pm 0.00002\,\pi\\ \qquad-\\ \qquad-\\ \qquad-\\ \qquad-\\ \qquad- \end{array}$ & $\begin{array}{ll} \;\;\;0.92386 &\pm 0.00352\\ \;\;\;0.00240 &\pm 0.00013\\ \;\;\;0.00471\,\pi &\pm 0.07067\,\pi\\ -\underline{0.04070\,\pi} &\pm 0.09111\,\pi\\ \;\;\;0.02131\,\pi &\pm 0.00151\,\pi\\ \;\;\;0.02161\,\pi &\pm 0.00174\,\pi\\ -0.02377\,\pi &\pm 0.00405\,\pi\\ \;\;\;0.03751 &\pm 0.00629\\ \;\;\;0.03800 &\pm 0.00628\\ \qquad-\\ \qquad-\\ \qquad- \end{array}$ & $\begin{array}{ll} \;\;\;0.92707 &\pm 0.00238\\ \;\;\;0.00006 &\pm 0.00147\\ -0.01176\,\pi &\pm 0.05021\,\pi\\ -\underline{0.06065\,\pi} &\pm 0.07914\,\pi\\ \;\;\;0.01876\,\pi &\pm 0.00183\,\pi\\ \;\;\;0.01876\,\pi &\pm 0.00183\,\pi\\ -0.02842\,\pi &\pm 0.00646\,\pi\\ \;\;\;0.01340 &\pm 0.00759\\ \;\;\;0.02078 &\pm 0.00539\\ \;\;\;0.06064 &\pm 0.00880\\ \;\;\;0.01920 &\pm 0.00587\\ \;\;\;0.03955 &\pm 0.01229\\ \end{array}$ \\
\noalign{\smallskip}
\hline
\noalign{\smallskip}
$\left\langle X\right\rangle$ & $\begin{array}{l} \eta\\ \varepsilon\\ \theta_1\\ \theta_2\\ \theta_3\\ \theta_4\\ \theta_5\\ \beta_1\\ \beta_2\\ \beta_3\\ \beta_4\\ \beta_5 \end{array}$ & $\begin{array}{ll} \;\;\;0.92975 &\pm 0.00020\\ \;\;\;0.03707 &\pm 0.00013\\ \qquad-\\ \qquad-\\ \qquad-\\ \qquad-\\ \qquad-\\ \qquad-\\ \qquad-\\ \qquad-\\ \qquad-\\ \qquad- \end{array}$ & $\begin{array}{ll} \;\;\;0.93039 &\pm 0.00009\\ \;\;\;0.03700 &\pm 0.00005\\ -0.00525\,\pi &\pm 0.00330\,\pi\\ -0.00573\,\pi &\pm 0.00330\,\pi\\ \;\;\;\underline{0.05396\,\pi} &\pm 0.07693\,\pi\\ \;\;\;\underline{0.02049\,\pi} &\pm 0.06694\,\pi\\ -0.00082\,\pi &\pm 0.00004\,\pi\\ \qquad-\\ \qquad-\\ \qquad-\\ \qquad-\\ \qquad- \end{array}$ & $\begin{array}{ll} \;\;\;0.94163 &\pm 0.00077\\ \;\;\;0.03706 &\pm 0.00005\\ -0.01759\,\pi &\pm 0.00056\,\pi\\ -0.01707\,\pi &\pm 0.00099\,\pi\\ \;\;\;\underline{0.10000\,\pi} &\pm 0.00000\,\pi\\ -\underline{0.01281\,\pi} &\pm 0.07203\,\pi\\ -0.05063\,\pi &\pm 0.00196\,\pi\\ \;\;\;0.17064 &\pm 0.00658\\ -0.01194 &\pm 0.00029\\ \qquad-\\ \qquad-\\ \qquad- \end{array}$ & $\begin{array}{ll} \;\;\;0.94054 &\pm 0.00170\\ \;\;\;0.03706 &\pm 0.00005\\ -0.00769\,\pi &\pm 0.00142\,\pi\\ -0.00767\,\pi &\pm 0.00142\,\pi\\ \;\;\;0.00961\,\pi &\pm 0.01214\,\pi\\ -\underline{0.00520\,\pi} &\pm 0.08215\,\pi\\ -0.00676\,\pi &\pm 0.00396\,\pi\\ \;\;\;0.10534 &\pm 0.02135\\ -0.03753 &\pm 0.01307\\ \;\;\;0.07033 &\pm 0.07038\\ -0.01431 &\pm 0.01384\\ -0.03618 &\pm 0.01304\\ \end{array}$ \\
\noalign{\smallskip}
\hline
\noalign{\smallskip}
$\left\langle X_0\right\rangle$ & $\begin{array}{l} \eta\\ \varepsilon\\ \theta_1\\ \theta_2\\ \theta_3\\ \theta_4\\ \theta_5\\ \beta_1\\ \beta_2\\ \beta_3\\ \beta_4\\ \beta_5 \end{array}$ & $\begin{array}{ll} \;\;\;0.91745 &\pm 0.00015\\ \;\;\;0.03394 &\pm 0.00008\\ \qquad-\\ \qquad-\\ \qquad-\\ \qquad-\\ \qquad-\\ \qquad-\\ \qquad-\\ \qquad-\\ \qquad-\\ \qquad- \end{array}$ & $\begin{array}{ll} \;\;\;0.91740 &\pm 0.00018\\ \;\;\;0.03360 &\pm 0.00007\\ -0.00331\,\pi &\pm 0.00002\,\pi\\ -0.00331\,\pi &\pm 0.00002\,\pi\\ \;\;\;\underline{0.00645\,\pi} &\pm 0.09506\,\pi\\ \;\;\;\underline{0.01769\,\pi} &\pm 0.07071\,\pi\\ -0.00331\,\pi &\pm 0.00002\,\pi\\ \qquad-\\ \qquad-\\ \qquad-\\ \qquad-\\ \qquad- \end{array}$ & $\begin{array}{ll} \;\;\;0.91900 &\pm 0.00011\\ \;\;\;0.03356 &\pm 0.00007\\ \;\;\;0.05805\,\pi &\pm 0.00009\,\pi\\ \;\;\;0.05805\,\pi &\pm 0.00010\,\pi\\ \;\;\;\underline{0.10000\,\pi} &\pm 0.00000\,\pi\\ -\underline{0.01414\,\pi} &\pm 0.08456\,\pi\\ -\underline{0.10000\,\pi} &\pm 0.00000\,\pi\\ -0.02477 &\pm 0.00063\\ -0.05963 &\pm 0.00029\\ \qquad-\\ \qquad-\\ \qquad- \end{array}$ & $\begin{array}{ll} \;\;\;0.94262 &\pm 0.00014\\ \;\;\;0.03237 &\pm 0.00005\\ \;\;\;0.00326\,\pi &\pm 0.00171\,\pi\\ \;\;\;0.00327\,\pi &\pm 0.00172\,\pi\\ -0.01545\,\pi &\pm 0.02125\,\pi\\ \;\;\;\underline{0.00289\,\pi} &\pm 0.09722\,\pi\\ \;\;\;0.00512\,\pi &\pm 0.00443\,\pi\\ -0.00394 &\pm 0.02631\\ -0.02698 &\pm 0.02227\\ \;\;\;0.23815 &\pm 0.00155\\ -0.06506 &\pm 0.03454\\ \;\;\;0.02333 &\pm 0.03759\\ \end{array}$ \\
\noalign{\smallskip}
\hline
\noalign{\smallskip}
$\left\langle X_1\right\rangle$ & $\begin{array}{l} \eta\\ \varepsilon\\ \theta_1\\ \theta_2\\ \theta_3\\ \theta_4\\ \theta_5\\ \beta_1\\ \beta_2\\ \beta_3\\ \beta_4\\ \beta_5 \end{array}$ & $\begin{array}{ll} \;\;\;0.93767 &\pm 0.00006\\ \;\;\;0.03391 &\pm 0.00004\\ \qquad-\\ \qquad-\\ \qquad-\\ \qquad-\\ \qquad-\\ \qquad-\\ \qquad-\\ \qquad-\\ \qquad-\\ \qquad- \end{array}$ & $\begin{array}{ll} \;\;\;0.93765 &\pm 0.00005\\ \;\;\;0.03379 &\pm 0.00006\\ -0.00137\,\pi &\pm 0.00047\,\pi\\ -0.00116\,\pi &\pm 0.00030\,\pi\\ -0.02542\,\pi &\pm 0.06461\,\pi\\ \;\;\;\underline{0.00000\,\pi} &\pm 0.09982\,\pi\\ \;\;\;0.00106\,\pi &\pm 0.00030\,\pi\\ \qquad-\\ \qquad-\\ \qquad-\\ \qquad-\\ \qquad- \end{array}$ & $\begin{array}{ll} \;\;\;0.93826 &\pm 0.00009\\ \;\;\;0.03382 &\pm 0.00005\\ \;\;\;0.05483\,\pi &\pm 0.00050\,\pi\\ \;\;\;0.05485\,\pi &\pm 0.00047\,\pi\\ -0.03627\,\pi &\pm 0.00287\,\pi\\ -\underline{0.03084\,\pi} &\pm 0.07377\,\pi\\ \;\;\;\underline{0.10000\,\pi} &\pm 0.00000\,\pi\\ -0.00272 &\pm 0.00027\\ \;\;\;0.04713 &\pm 0.00304\\ \qquad-\\ \qquad-\\ \qquad- \end{array}$ & $\begin{array}{ll} \;\;\;0.94136 &\pm 0.00007\\ \;\;\;0.03379 &\pm 0.00006\\ \;\;\;0.00477\,\pi &\pm 0.00164\,\pi\\ \;\;\;0.00477\,\pi &\pm 0.00164\,\pi\\ \;\;\;0.01919\,\pi &\pm 0.00582\,\pi\\ \;\;\;0.04070\,\pi &\pm 0.09111\,\pi\\ -\underline{0.00225\,\pi} &\pm 0.00134\,\pi\\ -0.01068 &\pm 0.00611\\ \;\;\;0.00853 &\pm 0.00825\\ \;\;\;0.06933 &\pm 0.00161\\ \;\;\;0.07775 &\pm 0.01049\\ -0.02379 &\pm 0.00541\\ \end{array}$ \\
\noalign{\smallskip}
\hline
\noalign{\smallskip}
${\cal D}_{m}$ & $\begin{array}{l} \eta\\ \varepsilon\\ \theta_1\\ \theta_2\\ \theta_3\\ \theta_4\\ \theta_5\\ \beta_1\\ \beta_2\\ \beta_3\\ \beta_4\\ \beta_5 \end{array}$ & $\begin{array}{ll} \qquad-\\ -0.00370 &\pm 0.00004\\ \qquad-\\ \qquad-\\ \qquad-\\ \qquad-\\ \qquad-\\ \qquad-\\ \qquad-\\ \qquad-\\ \qquad-\\ \qquad- \end{array}$ & $\begin{array}{ll} \;\;\;\underline{1.00000} &\pm 0.00000\\ -0.00370 &\pm 0.00007\\ \;\;\;0.00425\,\pi &\pm 0.07391\,\pi\\ -0.02309\,\pi &\pm 0.06652\,\pi\\ \;\;\;0.03880\,\pi &\pm 0.07545\,\pi\\ -0.00000\,\pi &\pm 0.00000\,\pi\\ -0.03308\,\pi &\pm 0.07253\,\pi\\ \qquad-\\ \qquad-\\ \qquad-\\ \qquad-\\ \qquad- \end{array}$ & $\begin{array}{ll} \;\;\;\underline{0.99999} &\pm 0.00002\\ \;\;\;0.01448 &\pm 0.00224\\ -\underline{0.01039\,\pi} &\pm 0.07781\,\pi\\ -\underline{0.00139\,\pi} &\pm 0.08569\,\pi\\ \;\;\;\underline{0.09986\,\pi} &\pm 0.00032\,\pi\\ -\underline{0.00591\,\pi} &\pm 0.06771\,\pi\\ -\underline{0.03750\,\pi} &\pm 0.08945\,\pi\\ -0.01573 &\pm 0.00234\\ -0.00364 &\pm 0.00140\\ \qquad-\\ \qquad-\\ \qquad- \end{array}$ & $\begin{array}{ll} \;\;\;0.98819 &\pm 0.03324\\ -0.32232 &\pm 0.01093\\ -0.00380\,\pi &\pm 0.06965\,\pi\\ \;\;\;\underline{0.00209\,\pi} &\pm 0.07158\,\pi\\ \;\;\;\underline{0.06215\,\pi} &\pm 0.04653\,\pi\\ -0.01688\,\pi &\pm 0.06911\,\pi\\ -\underline{0.09324\,\pi} &\pm 0.01471\,\pi\\ \;\;\;0.12234 &\pm 0.04425\\ -0.37500 &\pm 0.04456\\ \;\;\;0.05512 &\pm 0.01691\\ -0.10399 &\pm 0.04620\\ -0.39596 &\pm 0.05204 \end{array}$ \\
\noalign{\smallskip}
\hline
\hline
\end{tabularx}
\caption{Optimal model parameters obtained with different single- and two-qubit gate models, evaluated by resorting to the ten different fits used in tenfold cross-validation. Underlined values are unstable cases in which some of the fits returned the value at one of the imposed bounds, occurring at only one of them in all ten different splits when precision errors indicate zero.}
\label{tab:all-fits-params}
\end{minipage}
\end{table}

\FloatBarrier

\bibliography{biblio}

\begin{thebibliography}{74}%
\makeatletter
\providecommand \@ifxundefined [1]{%
 \@ifx{#1\undefined}
}%
\providecommand \@ifnum [1]{%
 \ifnum #1\expandafter \@firstoftwo
 \else \expandafter \@secondoftwo
 \fi
}%
\providecommand \@ifx [1]{%
 \ifx #1\expandafter \@firstoftwo
 \else \expandafter \@secondoftwo
 \fi
}%
\providecommand \natexlab [1]{#1}%
\providecommand \enquote  [1]{``#1''}%
\providecommand \bibnamefont  [1]{#1}%
\providecommand \bibfnamefont [1]{#1}%
\providecommand \citenamefont [1]{#1}%
\providecommand \href@noop [0]{\@secondoftwo}%
\providecommand \href [0]{\begingroup \@sanitize@url \@href}%
\providecommand \@href[1]{\@@startlink{#1}\@@href}%
\providecommand \@@href[1]{\endgroup#1\@@endlink}%
\providecommand \@sanitize@url [0]{\catcode `\\12\catcode `\$12\catcode
  `\&12\catcode `\#12\catcode `\^12\catcode `\_12\catcode `\%12\relax}%
\providecommand \@@startlink[1]{}%
\providecommand \@@endlink[0]{}%
\providecommand \url  [0]{\begingroup\@sanitize@url \@url }%
\providecommand \@url [1]{\endgroup\@href {#1}{\urlprefix }}%
\providecommand \urlprefix  [0]{URL }%
\providecommand \Eprint [0]{\href }%
\providecommand \doibase [0]{https://doi.org/}%
\providecommand \selectlanguage [0]{\@gobble}%
\providecommand \bibinfo  [0]{\@secondoftwo}%
\providecommand \bibfield  [0]{\@secondoftwo}%
\providecommand \translation [1]{[#1]}%
\providecommand \BibitemOpen [0]{}%
\providecommand \bibitemStop [0]{}%
\providecommand \bibitemNoStop [0]{.\EOS\space}%
\providecommand \EOS [0]{\spacefactor3000\relax}%
\providecommand \BibitemShut  [1]{\csname bibitem#1\endcsname}%
\let\auto@bib@innerbib\@empty
%</preamble>
\bibitem [{\citenamefont {Bohr}(1928)}]{bohr1928quantum}%
  \BibitemOpen
  \bibfield  {author} {\bibinfo {author} {\bibfnamefont {N.}~\bibnamefont
  {Bohr}},\ }\bibfield  {title} {\bibinfo {title} {The quantum postulate and
  the recent development of atomic theory},\ }\href@noop {} {\bibfield
  {journal} {\bibinfo  {journal} {Nature}\ }\textbf {\bibinfo {volume} {121}},\
  \bibinfo {pages} {580} (\bibinfo {year} {1928})}\BibitemShut {NoStop}%
\bibitem [{\citenamefont {De~Broglie}(1924)}]{debroglie1924recherches}%
  \BibitemOpen
  \bibfield  {author} {\bibinfo {author} {\bibfnamefont {L.}~\bibnamefont
  {De~Broglie}},\ }\emph {\bibinfo {title} {Recherches sur la th{\'e}orie des
  quanta}},\ \href@noop {} {Ph.D. thesis},\ \bibinfo  {school}
  {Migration-universit{\'e} en cours d'affectation} (\bibinfo {year}
  {1924})\BibitemShut {NoStop}%
\bibitem [{\citenamefont {Heisenberg}(1927)}]{heisenberg1927uber}%
  \BibitemOpen
  \bibfield  {author} {\bibinfo {author} {\bibfnamefont {W.}~\bibnamefont
  {Heisenberg}},\ }\bibfield  {title} {\bibinfo {title} {{\"U}ber den
  anschaulichen inhalt der quantentheoretischen kinematik und mechanik},\
  }\href@noop {} {\bibfield  {journal} {\bibinfo  {journal} {Zeitschrift
  f{\"u}r Physik}\ }\textbf {\bibinfo {volume} {43}},\ \bibinfo {pages} {172}
  (\bibinfo {year} {1927})}\BibitemShut {NoStop}%
\bibitem [{\citenamefont {Bohr}(1949)}]{bohr1949discussion}%
  \BibitemOpen
  \bibfield  {author} {\bibinfo {author} {\bibfnamefont {N.}~\bibnamefont
  {Bohr}},\ }\bibinfo {title} {Albert einstein: Philosopher-scientist}\
  (\bibinfo  {publisher} {Library of Living Philosophers},\ \bibinfo {address}
  {Evanston, Illinois},\ \bibinfo {year} {1949})\ Chap.\ \bibinfo {chapter} {7
  - Discussion with Einstein on Epistemological Problems in Atomic Physics},
  pp.\ \bibinfo {pages} {201--241}\BibitemShut {NoStop}%
\bibitem [{\citenamefont {Heisenberg}(1930)}]{heisenberg1930physical}%
  \BibitemOpen
  \bibfield  {author} {\bibinfo {author} {\bibfnamefont {W.}~\bibnamefont
  {Heisenberg}},\ }\href@noop {} {\emph {\bibinfo {title} {The Physical
  Principles of the Quantum Theory}}}\ (\bibinfo  {publisher} {University of
  Chicago Press},\ \bibinfo {year} {1930})\BibitemShut {NoStop}%
\bibitem [{\citenamefont {Feynman}\ \emph {et~al.}(1965)\citenamefont
  {Feynman}, \citenamefont {Leighton},\ and\ \citenamefont
  {Sands}}]{feynman1965feynman}%
  \BibitemOpen
  \bibfield  {author} {\bibinfo {author} {\bibfnamefont {R.}~\bibnamefont
  {Feynman}}, \bibinfo {author} {\bibfnamefont {R.}~\bibnamefont {Leighton}},\
  and\ \bibinfo {author} {\bibfnamefont {M.}~\bibnamefont {Sands}},\
  }\href@noop {} {\emph {\bibinfo {title} {The Feynman Lectures on Physics,
  Vol. III}}}\ (\bibinfo  {publisher} {Addison-Wesley},\ \bibinfo {year}
  {1965})\BibitemShut {NoStop}%
\bibitem [{\citenamefont {Wheeler}(1978)}]{wheeler1978past}%
  \BibitemOpen
  \bibfield  {author} {\bibinfo {author} {\bibfnamefont {J.~A.}\ \bibnamefont
  {Wheeler}},\ }\bibfield  {title} {\bibinfo {title} {The “past” and the
  “delayed-choice” double-slit experiment},\ }in\ \href@noop {} {\emph
  {\bibinfo {booktitle} {Mathematical Foundations of Quantum Theory}}},\
  \bibinfo {editor} {edited by\ \bibinfo {editor} {\bibfnamefont
  {A.}~\bibnamefont {Marlow}}}\ (\bibinfo  {publisher} {Academic Press},\
  \bibinfo {year} {1978})\ pp.\ \bibinfo {pages} {9 -- 48}\BibitemShut
  {NoStop}%
\bibitem [{\citenamefont {Wootters}\ and\ \citenamefont
  {Zurek}(1979)}]{wootters1979complementarity}%
  \BibitemOpen
  \bibfield  {author} {\bibinfo {author} {\bibfnamefont {W.~K.}\ \bibnamefont
  {Wootters}}\ and\ \bibinfo {author} {\bibfnamefont {W.~H.}\ \bibnamefont
  {Zurek}},\ }\bibfield  {title} {\bibinfo {title} {Complementarity in the
  double-slit experiment: Quantum nonseparability and a quantitative statement
  of bohr's principle},\ }\href@noop {} {\bibfield  {journal} {\bibinfo
  {journal} {Physical Review D}\ }\textbf {\bibinfo {volume} {19}},\ \bibinfo
  {pages} {473} (\bibinfo {year} {1979})}\BibitemShut {NoStop}%
\bibitem [{\citenamefont {Scully}\ \emph {et~al.}(1989)\citenamefont {Scully},
  \citenamefont {Englert},\ and\ \citenamefont {Schwinger}}]{scully1989spin}%
  \BibitemOpen
  \bibfield  {author} {\bibinfo {author} {\bibfnamefont {M.~O.}\ \bibnamefont
  {Scully}}, \bibinfo {author} {\bibfnamefont {B.-G.}\ \bibnamefont
  {Englert}},\ and\ \bibinfo {author} {\bibfnamefont {J.}~\bibnamefont
  {Schwinger}},\ }\bibfield  {title} {\bibinfo {title} {Spin coherence and
  humpty-dumpty. iii. the effects of observation},\ }\href@noop {} {\bibfield
  {journal} {\bibinfo  {journal} {Phys. Rev. A}\ }\textbf {\bibinfo {volume}
  {40}},\ \bibinfo {pages} {1775} (\bibinfo {year} {1989})}\BibitemShut
  {NoStop}%
\bibitem [{\citenamefont {Scully}\ and\ \citenamefont
  {Walther}(1989)}]{scully1989quantum}%
  \BibitemOpen
  \bibfield  {author} {\bibinfo {author} {\bibfnamefont {M.~O.}\ \bibnamefont
  {Scully}}\ and\ \bibinfo {author} {\bibfnamefont {H.}~\bibnamefont
  {Walther}},\ }\bibfield  {title} {\bibinfo {title} {Quantum optical test of
  observation and complementarity in quantum mechanics},\ }\href@noop {}
  {\bibfield  {journal} {\bibinfo  {journal} {Phys. Rev. A}\ }\textbf {\bibinfo
  {volume} {39}},\ \bibinfo {pages} {5229} (\bibinfo {year}
  {1989})}\BibitemShut {NoStop}%
\bibitem [{\citenamefont {Deutsch}(1983)}]{deutsch1983uncertainty}%
  \BibitemOpen
  \bibfield  {author} {\bibinfo {author} {\bibfnamefont {D.}~\bibnamefont
  {Deutsch}},\ }\bibfield  {title} {\bibinfo {title} {Uncertainty in quantum
  measurements},\ }\href@noop {} {\bibfield  {journal} {\bibinfo  {journal}
  {Phys. Rev. Lett.}\ }\textbf {\bibinfo {volume} {50}},\ \bibinfo {pages}
  {631} (\bibinfo {year} {1983})}\BibitemShut {NoStop}%
\bibitem [{\citenamefont {Mittelstaedt}\ \emph {et~al.}(1987)\citenamefont
  {Mittelstaedt}, \citenamefont {Prieur},\ and\ \citenamefont
  {Schieder}}]{mittelstaedt1987unsharp}%
  \BibitemOpen
  \bibfield  {author} {\bibinfo {author} {\bibfnamefont {P.}~\bibnamefont
  {Mittelstaedt}}, \bibinfo {author} {\bibfnamefont {A.}~\bibnamefont
  {Prieur}},\ and\ \bibinfo {author} {\bibfnamefont {R.}~\bibnamefont
  {Schieder}},\ }\bibfield  {title} {\bibinfo {title} {Unsharp particle-wave
  duality in a photon split-beam experiment},\ }\href@noop {} {\bibfield
  {journal} {\bibinfo  {journal} {Foundations of Physics}\ }\textbf {\bibinfo
  {volume} {17}},\ \bibinfo {pages} {891} (\bibinfo {year} {1987})}\BibitemShut
  {NoStop}%
\bibitem [{\citenamefont {Greenberger}\ and\ \citenamefont
  {Yasin}(1988)}]{greenberger1988simultaneous}%
  \BibitemOpen
  \bibfield  {author} {\bibinfo {author} {\bibfnamefont {D.~M.}\ \bibnamefont
  {Greenberger}}\ and\ \bibinfo {author} {\bibfnamefont {A.}~\bibnamefont
  {Yasin}},\ }\bibfield  {title} {\bibinfo {title} {Simultaneous wave and
  particle knowledge in a neutron interferometer},\ }\href@noop {} {\bibfield
  {journal} {\bibinfo  {journal} {Physics Letters A}\ }\textbf {\bibinfo
  {volume} {128}},\ \bibinfo {pages} {391} (\bibinfo {year}
  {1988})}\BibitemShut {NoStop}%
\bibitem [{\citenamefont {Jaeger}\ \emph {et~al.}(1995)\citenamefont {Jaeger},
  \citenamefont {Shimony},\ and\ \citenamefont {Vaidman}}]{jaeger1995two}%
  \BibitemOpen
  \bibfield  {author} {\bibinfo {author} {\bibfnamefont {G.}~\bibnamefont
  {Jaeger}}, \bibinfo {author} {\bibfnamefont {A.}~\bibnamefont {Shimony}},\
  and\ \bibinfo {author} {\bibfnamefont {L.}~\bibnamefont {Vaidman}},\
  }\bibfield  {title} {\bibinfo {title} {Two interferometric
  complementarities},\ }\href@noop {} {\bibfield  {journal} {\bibinfo
  {journal} {Physical Review A}\ }\textbf {\bibinfo {volume} {51}},\ \bibinfo
  {pages} {54} (\bibinfo {year} {1995})}\BibitemShut {NoStop}%
\bibitem [{\citenamefont {Englert}(1996)}]{englert1996fringe}%
  \BibitemOpen
  \bibfield  {author} {\bibinfo {author} {\bibfnamefont {B.-G.}\ \bibnamefont
  {Englert}},\ }\bibfield  {title} {\bibinfo {title} {Fringe visibility and
  which-way information: An inequality},\ }\href@noop {} {\bibfield  {journal}
  {\bibinfo  {journal} {Phys. Rev. Lett.}\ }\textbf {\bibinfo {volume} {77}},\
  \bibinfo {pages} {2154} (\bibinfo {year} {1996})}\BibitemShut {NoStop}%
\bibitem [{\citenamefont {Bj\"ork}\ and\ \citenamefont
  {Karlsson}(1998)}]{bjork1998complementarity}%
  \BibitemOpen
  \bibfield  {author} {\bibinfo {author} {\bibfnamefont {G.}~\bibnamefont
  {Bj\"ork}}\ and\ \bibinfo {author} {\bibfnamefont {A.}~\bibnamefont
  {Karlsson}},\ }\bibfield  {title} {\bibinfo {title} {Complementarity and
  quantum erasure in welcher weg experiments},\ }\href@noop {} {\bibfield
  {journal} {\bibinfo  {journal} {Phys. Rev. A}\ }\textbf {\bibinfo {volume}
  {58}},\ \bibinfo {pages} {3477} (\bibinfo {year} {1998})}\BibitemShut
  {NoStop}%
\bibitem [{\citenamefont {D\"urr}(2001)}]{durr2001quantitative}%
  \BibitemOpen
  \bibfield  {author} {\bibinfo {author} {\bibfnamefont {S.}~\bibnamefont
  {D\"urr}},\ }\bibfield  {title} {\bibinfo {title} {Quantitative wave-particle
  duality in multibeam interferometers},\ }\href@noop {} {\bibfield  {journal}
  {\bibinfo  {journal} {Phys. Rev. A}\ }\textbf {\bibinfo {volume} {64}},\
  \bibinfo {pages} {042113} (\bibinfo {year} {2001})}\BibitemShut {NoStop}%
\bibitem [{\citenamefont {Jakob}\ and\ \citenamefont
  {Bergou}(2007)}]{jakob2007complementarity}%
  \BibitemOpen
  \bibfield  {author} {\bibinfo {author} {\bibfnamefont {M.}~\bibnamefont
  {Jakob}}\ and\ \bibinfo {author} {\bibfnamefont {J.~A.}\ \bibnamefont
  {Bergou}},\ }\bibfield  {title} {\bibinfo {title} {Complementarity and
  entanglement in bipartite qudit systems},\ }\href@noop {} {\bibfield
  {journal} {\bibinfo  {journal} {Phys. Rev. A}\ }\textbf {\bibinfo {volume}
  {76}},\ \bibinfo {pages} {052107} (\bibinfo {year} {2007})}\BibitemShut
  {NoStop}%
\bibitem [{\citenamefont {Jakob}\ and\ \citenamefont
  {Bergou}(2010)}]{jakob2010quantitative}%
  \BibitemOpen
  \bibfield  {author} {\bibinfo {author} {\bibfnamefont {M.}~\bibnamefont
  {Jakob}}\ and\ \bibinfo {author} {\bibfnamefont {J.~A.}\ \bibnamefont
  {Bergou}},\ }\bibfield  {title} {\bibinfo {title} {Quantitative
  complementarity relations in bipartite systems: Entanglement as a physical
  reality},\ }\href@noop {} {\bibfield  {journal} {\bibinfo  {journal} {Optics
  Communications}\ }\textbf {\bibinfo {volume} {283}},\ \bibinfo {pages} {827 }
  (\bibinfo {year} {2010})}\BibitemShut {NoStop}%
\bibitem [{\citenamefont {Coles}\ \emph {et~al.}(2014)\citenamefont {Coles},
  \citenamefont {Kaniewski},\ and\ \citenamefont
  {Wehner}}]{coles2014equivalence}%
  \BibitemOpen
  \bibfield  {author} {\bibinfo {author} {\bibfnamefont {P.~J.}\ \bibnamefont
  {Coles}}, \bibinfo {author} {\bibfnamefont {J.}~\bibnamefont {Kaniewski}},\
  and\ \bibinfo {author} {\bibfnamefont {S.}~\bibnamefont {Wehner}},\
  }\bibfield  {title} {\bibinfo {title} {Equivalence of wave--particle duality
  to entropic uncertainty},\ }\href@noop {} {\bibfield  {journal} {\bibinfo
  {journal} {Nature communications}\ }\textbf {\bibinfo {volume} {5}},\
  \bibinfo {pages} {5814} (\bibinfo {year} {2014})}\BibitemShut {NoStop}%
\bibitem [{\citenamefont {Coles}(2016)}]{coles2016entropic}%
  \BibitemOpen
  \bibfield  {author} {\bibinfo {author} {\bibfnamefont {P.~J.}\ \bibnamefont
  {Coles}},\ }\bibfield  {title} {\bibinfo {title} {Entropic framework for
  wave-particle duality in multipath interferometers},\ }\href@noop {}
  {\bibfield  {journal} {\bibinfo  {journal} {Phys. Rev. A}\ }\textbf {\bibinfo
  {volume} {93}},\ \bibinfo {pages} {062111} (\bibinfo {year}
  {2016})}\BibitemShut {NoStop}%
\bibitem [{\citenamefont {Angelo}\ and\ \citenamefont
  {Ribeiro}(2015)}]{angelo2015wave}%
  \BibitemOpen
  \bibfield  {author} {\bibinfo {author} {\bibfnamefont {R.~M.}\ \bibnamefont
  {Angelo}}\ and\ \bibinfo {author} {\bibfnamefont {A.~D.}\ \bibnamefont
  {Ribeiro}},\ }\bibfield  {title} {\bibinfo {title} {Wave--particle duality:
  An information-based approach},\ }\href@noop {} {\bibfield  {journal}
  {\bibinfo  {journal} {Foundations of Physics}\ }\textbf {\bibinfo {volume}
  {45}},\ \bibinfo {pages} {1407} (\bibinfo {year} {2015})}\BibitemShut
  {NoStop}%
\bibitem [{\citenamefont {Bagan}\ \emph {et~al.}(2016)\citenamefont {Bagan},
  \citenamefont {Bergou}, \citenamefont {Cottrell},\ and\ \citenamefont
  {Hillery}}]{bagan2016relations}%
  \BibitemOpen
  \bibfield  {author} {\bibinfo {author} {\bibfnamefont {E.}~\bibnamefont
  {Bagan}}, \bibinfo {author} {\bibfnamefont {J.~A.}\ \bibnamefont {Bergou}},
  \bibinfo {author} {\bibfnamefont {S.~S.}\ \bibnamefont {Cottrell}},\ and\
  \bibinfo {author} {\bibfnamefont {M.}~\bibnamefont {Hillery}},\ }\bibfield
  {title} {\bibinfo {title} {Relations between coherence and path
  information},\ }\href@noop {} {\bibfield  {journal} {\bibinfo  {journal}
  {Phys. Rev. Lett.}\ }\textbf {\bibinfo {volume} {116}},\ \bibinfo {pages}
  {160406} (\bibinfo {year} {2016})}\BibitemShut {NoStop}%
\bibitem [{\citenamefont {D{\"u}rr}\ \emph {et~al.}(1998)\citenamefont
  {D{\"u}rr}, \citenamefont {Nonn},\ and\ \citenamefont
  {Rempe}}]{durr1998origin}%
  \BibitemOpen
  \bibfield  {author} {\bibinfo {author} {\bibfnamefont {S.}~\bibnamefont
  {D{\"u}rr}}, \bibinfo {author} {\bibfnamefont {T.}~\bibnamefont {Nonn}},\
  and\ \bibinfo {author} {\bibfnamefont {G.}~\bibnamefont {Rempe}},\ }\bibfield
   {title} {\bibinfo {title} {Origin of quantum-mechanical complementarity
  probed by a `which-way' experiment in an atom interferometer},\ }\href@noop
  {} {\bibfield  {journal} {\bibinfo  {journal} {Nature}\ }\textbf {\bibinfo
  {volume} {395}},\ \bibinfo {pages} {33} (\bibinfo {year} {1998})}\BibitemShut
  {NoStop}%
\bibitem [{\citenamefont {Bertet}\ \emph {et~al.}(2001)\citenamefont {Bertet},
  \citenamefont {Osnaghi}, \citenamefont {Rauschenbeutel}, \citenamefont
  {Nogues}, \citenamefont {Auffeves}, \citenamefont {Brune}, \citenamefont
  {Raimond},\ and\ \citenamefont {Haroche}}]{bertet2001complementarity}%
  \BibitemOpen
  \bibfield  {author} {\bibinfo {author} {\bibfnamefont {P.}~\bibnamefont
  {Bertet}}, \bibinfo {author} {\bibfnamefont {S.}~\bibnamefont {Osnaghi}},
  \bibinfo {author} {\bibfnamefont {A.}~\bibnamefont {Rauschenbeutel}},
  \bibinfo {author} {\bibfnamefont {G.}~\bibnamefont {Nogues}}, \bibinfo
  {author} {\bibfnamefont {A.}~\bibnamefont {Auffeves}}, \bibinfo {author}
  {\bibfnamefont {M.}~\bibnamefont {Brune}}, \bibinfo {author} {\bibfnamefont
  {J.}~\bibnamefont {Raimond}},\ and\ \bibinfo {author} {\bibfnamefont
  {S.}~\bibnamefont {Haroche}},\ }\bibfield  {title} {\bibinfo {title} {A
  complementarity experiment with an interferometer at the quantum--classical
  boundary},\ }\href@noop {} {\bibfield  {journal} {\bibinfo  {journal}
  {Nature}\ }\textbf {\bibinfo {volume} {411}},\ \bibinfo {pages} {166}
  (\bibinfo {year} {2001})}\BibitemShut {NoStop}%
\bibitem [{\citenamefont {Herzog}\ \emph {et~al.}(1995)\citenamefont {Herzog},
  \citenamefont {Kwiat}, \citenamefont {Weinfurter},\ and\ \citenamefont
  {Zeilinger}}]{herzog1995complementarity}%
  \BibitemOpen
  \bibfield  {author} {\bibinfo {author} {\bibfnamefont {T.~J.}\ \bibnamefont
  {Herzog}}, \bibinfo {author} {\bibfnamefont {P.~G.}\ \bibnamefont {Kwiat}},
  \bibinfo {author} {\bibfnamefont {H.}~\bibnamefont {Weinfurter}},\ and\
  \bibinfo {author} {\bibfnamefont {A.}~\bibnamefont {Zeilinger}},\ }\bibfield
  {title} {\bibinfo {title} {Complementarity and the quantum eraser},\
  }\href@noop {} {\bibfield  {journal} {\bibinfo  {journal} {Phys. Rev. Lett.}\
  }\textbf {\bibinfo {volume} {75}},\ \bibinfo {pages} {3034} (\bibinfo {year}
  {1995})}\BibitemShut {NoStop}%
\bibitem [{\citenamefont {Kim}\ \emph {et~al.}(2000)\citenamefont {Kim},
  \citenamefont {Yu}, \citenamefont {Kulik}, \citenamefont {Shih},\ and\
  \citenamefont {Scully}}]{kim2000delayed}%
  \BibitemOpen
  \bibfield  {author} {\bibinfo {author} {\bibfnamefont {Y.-H.}\ \bibnamefont
  {Kim}}, \bibinfo {author} {\bibfnamefont {R.}~\bibnamefont {Yu}}, \bibinfo
  {author} {\bibfnamefont {S.~P.}\ \bibnamefont {Kulik}}, \bibinfo {author}
  {\bibfnamefont {Y.}~\bibnamefont {Shih}},\ and\ \bibinfo {author}
  {\bibfnamefont {M.~O.}\ \bibnamefont {Scully}},\ }\bibfield  {title}
  {\bibinfo {title} {Delayed ``choice'' quantum eraser},\ }\href@noop {}
  {\bibfield  {journal} {\bibinfo  {journal} {Phys. Rev. Lett.}\ }\textbf
  {\bibinfo {volume} {84}},\ \bibinfo {pages} {1} (\bibinfo {year}
  {2000})}\BibitemShut {NoStop}%
\bibitem [{\citenamefont {Walborn}\ \emph {et~al.}(2002)\citenamefont
  {Walborn}, \citenamefont {Terra~Cunha}, \citenamefont {P\'adua},\ and\
  \citenamefont {Monken}}]{walborn2002doubleslit}%
  \BibitemOpen
  \bibfield  {author} {\bibinfo {author} {\bibfnamefont {S.~P.}\ \bibnamefont
  {Walborn}}, \bibinfo {author} {\bibfnamefont {M.~O.}\ \bibnamefont
  {Terra~Cunha}}, \bibinfo {author} {\bibfnamefont {S.}~\bibnamefont
  {P\'adua}},\ and\ \bibinfo {author} {\bibfnamefont {C.~H.}\ \bibnamefont
  {Monken}},\ }\bibfield  {title} {\bibinfo {title} {Double-slit quantum
  eraser},\ }\href@noop {} {\bibfield  {journal} {\bibinfo  {journal} {Phys.
  Rev. A}\ }\textbf {\bibinfo {volume} {65}},\ \bibinfo {pages} {033818}
  (\bibinfo {year} {2002})}\BibitemShut {NoStop}%
\bibitem [{\citenamefont {Neves}\ \emph {et~al.}(2009)\citenamefont {Neves},
  \citenamefont {Lima}, \citenamefont {Aguirre}, \citenamefont {Torres-Ruiz},
  \citenamefont {Saavedra},\ and\ \citenamefont {Delgado}}]{neves2009control}%
  \BibitemOpen
  \bibfield  {author} {\bibinfo {author} {\bibfnamefont {L.}~\bibnamefont
  {Neves}}, \bibinfo {author} {\bibfnamefont {G.}~\bibnamefont {Lima}},
  \bibinfo {author} {\bibfnamefont {J.}~\bibnamefont {Aguirre}}, \bibinfo
  {author} {\bibfnamefont {F.~A.}\ \bibnamefont {Torres-Ruiz}}, \bibinfo
  {author} {\bibfnamefont {C.}~\bibnamefont {Saavedra}},\ and\ \bibinfo
  {author} {\bibfnamefont {A.}~\bibnamefont {Delgado}},\ }\bibfield  {title}
  {\bibinfo {title} {Control of quantum interference in the quantum eraser},\
  }\href@noop {} {\bibfield  {journal} {\bibinfo  {journal} {New Journal of
  Physics}\ }\textbf {\bibinfo {volume} {11}},\ \bibinfo {pages} {073035}
  (\bibinfo {year} {2009})}\BibitemShut {NoStop}%
\bibitem [{\citenamefont {Peruzzo}\ \emph {et~al.}(2012)\citenamefont
  {Peruzzo}, \citenamefont {Shadbolt}, \citenamefont {Brunner}, \citenamefont
  {Popescu},\ and\ \citenamefont
  {O{\textquoteright}Brien}}]{peruzzo2012aquantum}%
  \BibitemOpen
  \bibfield  {author} {\bibinfo {author} {\bibfnamefont {A.}~\bibnamefont
  {Peruzzo}}, \bibinfo {author} {\bibfnamefont {P.}~\bibnamefont {Shadbolt}},
  \bibinfo {author} {\bibfnamefont {N.}~\bibnamefont {Brunner}}, \bibinfo
  {author} {\bibfnamefont {S.}~\bibnamefont {Popescu}},\ and\ \bibinfo {author}
  {\bibfnamefont {J.~L.}\ \bibnamefont {O{\textquoteright}Brien}},\ }\bibfield
  {title} {\bibinfo {title} {A quantum delayed-choice experiment},\ }\href@noop
  {} {\bibfield  {journal} {\bibinfo  {journal} {Science}\ }\textbf {\bibinfo
  {volume} {338}},\ \bibinfo {pages} {634} (\bibinfo {year}
  {2012})}\BibitemShut {NoStop}%
\bibitem [{\citenamefont {Kaiser}\ \emph {et~al.}(2012)\citenamefont {Kaiser},
  \citenamefont {Coudreau}, \citenamefont {Milman}, \citenamefont {Ostrowsky},\
  and\ \citenamefont {Tanzilli}}]{kaiser2012entanglement}%
  \BibitemOpen
  \bibfield  {author} {\bibinfo {author} {\bibfnamefont {F.}~\bibnamefont
  {Kaiser}}, \bibinfo {author} {\bibfnamefont {T.}~\bibnamefont {Coudreau}},
  \bibinfo {author} {\bibfnamefont {P.}~\bibnamefont {Milman}}, \bibinfo
  {author} {\bibfnamefont {D.~B.}\ \bibnamefont {Ostrowsky}},\ and\ \bibinfo
  {author} {\bibfnamefont {S.}~\bibnamefont {Tanzilli}},\ }\bibfield  {title}
  {\bibinfo {title} {Entanglement-enabled delayed-choice experiment},\
  }\href@noop {} {\bibfield  {journal} {\bibinfo  {journal} {Science}\ }\textbf
  {\bibinfo {volume} {338}},\ \bibinfo {pages} {637} (\bibinfo {year}
  {2012})}\BibitemShut {NoStop}%
\bibitem [{\citenamefont {Buks}\ \emph {et~al.}(1998)\citenamefont {Buks},
  \citenamefont {Schuster}, \citenamefont {Heiblum}, \citenamefont {Mahalu},\
  and\ \citenamefont {Umansky}}]{buks1998dephasing}%
  \BibitemOpen
  \bibfield  {author} {\bibinfo {author} {\bibfnamefont {E.}~\bibnamefont
  {Buks}}, \bibinfo {author} {\bibfnamefont {R.}~\bibnamefont {Schuster}},
  \bibinfo {author} {\bibfnamefont {M.}~\bibnamefont {Heiblum}}, \bibinfo
  {author} {\bibfnamefont {D.}~\bibnamefont {Mahalu}},\ and\ \bibinfo {author}
  {\bibfnamefont {V.}~\bibnamefont {Umansky}},\ }\bibfield  {title} {\bibinfo
  {title} {Dephasing in electron interference by a `which-path' detector},\
  }\href@noop {} {\bibfield  {journal} {\bibinfo  {journal} {Nature}\ }\textbf
  {\bibinfo {volume} {391}},\ \bibinfo {pages} {871} (\bibinfo {year}
  {1998})}\BibitemShut {NoStop}%
\bibitem [{\citenamefont {Sprinzak}\ \emph {et~al.}(2000)\citenamefont
  {Sprinzak}, \citenamefont {Buks}, \citenamefont {Heiblum},\ and\
  \citenamefont {Shtrikman}}]{sprinzak2000controlled}%
  \BibitemOpen
  \bibfield  {author} {\bibinfo {author} {\bibfnamefont {D.}~\bibnamefont
  {Sprinzak}}, \bibinfo {author} {\bibfnamefont {E.}~\bibnamefont {Buks}},
  \bibinfo {author} {\bibfnamefont {M.}~\bibnamefont {Heiblum}},\ and\ \bibinfo
  {author} {\bibfnamefont {H.}~\bibnamefont {Shtrikman}},\ }\bibfield  {title}
  {\bibinfo {title} {Controlled dephasing of electrons via a phase sensitive
  detector},\ }\href@noop {} {\bibfield  {journal} {\bibinfo  {journal} {Phys.
  Rev. Lett.}\ }\textbf {\bibinfo {volume} {84}},\ \bibinfo {pages} {5820}
  (\bibinfo {year} {2000})}\BibitemShut {NoStop}%
\bibitem [{\citenamefont {Neder}\ \emph {et~al.}(2007)\citenamefont {Neder},
  \citenamefont {Heiblum}, \citenamefont {Mahalu},\ and\ \citenamefont
  {Umansky}}]{neder2007entanglement}%
  \BibitemOpen
  \bibfield  {author} {\bibinfo {author} {\bibfnamefont {I.}~\bibnamefont
  {Neder}}, \bibinfo {author} {\bibfnamefont {M.}~\bibnamefont {Heiblum}},
  \bibinfo {author} {\bibfnamefont {D.}~\bibnamefont {Mahalu}},\ and\ \bibinfo
  {author} {\bibfnamefont {V.}~\bibnamefont {Umansky}},\ }\bibfield  {title}
  {\bibinfo {title} {Entanglement, dephasing, and phase recovery via
  cross-correlation measurements of electrons},\ }\href@noop {} {\bibfield
  {journal} {\bibinfo  {journal} {Phys. Rev. Lett.}\ }\textbf {\bibinfo
  {volume} {98}},\ \bibinfo {pages} {036803} (\bibinfo {year}
  {2007})}\BibitemShut {NoStop}%
\bibitem [{\citenamefont {Teklemariam}\ \emph {et~al.}(2001)\citenamefont
  {Teklemariam}, \citenamefont {Fortunato}, \citenamefont {Pravia},
  \citenamefont {Havel},\ and\ \citenamefont {Cory}}]{teklemariam2001NMR}%
  \BibitemOpen
  \bibfield  {author} {\bibinfo {author} {\bibfnamefont {G.}~\bibnamefont
  {Teklemariam}}, \bibinfo {author} {\bibfnamefont {E.~M.}\ \bibnamefont
  {Fortunato}}, \bibinfo {author} {\bibfnamefont {M.~A.}\ \bibnamefont
  {Pravia}}, \bibinfo {author} {\bibfnamefont {T.~F.}\ \bibnamefont {Havel}},\
  and\ \bibinfo {author} {\bibfnamefont {D.~G.}\ \bibnamefont {Cory}},\
  }\bibfield  {title} {\bibinfo {title} {Nmr analog of the quantum
  disentanglement eraser},\ }\href@noop {} {\bibfield  {journal} {\bibinfo
  {journal} {Phys. Rev. Lett.}\ }\textbf {\bibinfo {volume} {86}},\ \bibinfo
  {pages} {5845} (\bibinfo {year} {2001})}\BibitemShut {NoStop}%
\bibitem [{\citenamefont {Teklemariam}\ \emph {et~al.}(2002)\citenamefont
  {Teklemariam}, \citenamefont {Fortunato}, \citenamefont {Pravia},
  \citenamefont {Sharf}, \citenamefont {Havel}, \citenamefont {Cory},
  \citenamefont {Bhattaharyya},\ and\ \citenamefont
  {Hou}}]{teklemariam2002quantum}%
  \BibitemOpen
  \bibfield  {author} {\bibinfo {author} {\bibfnamefont {G.}~\bibnamefont
  {Teklemariam}}, \bibinfo {author} {\bibfnamefont {E.~M.}\ \bibnamefont
  {Fortunato}}, \bibinfo {author} {\bibfnamefont {M.~A.}\ \bibnamefont
  {Pravia}}, \bibinfo {author} {\bibfnamefont {Y.}~\bibnamefont {Sharf}},
  \bibinfo {author} {\bibfnamefont {T.~F.}\ \bibnamefont {Havel}}, \bibinfo
  {author} {\bibfnamefont {D.~G.}\ \bibnamefont {Cory}}, \bibinfo {author}
  {\bibfnamefont {A.}~\bibnamefont {Bhattaharyya}},\ and\ \bibinfo {author}
  {\bibfnamefont {J.}~\bibnamefont {Hou}},\ }\bibfield  {title} {\bibinfo
  {title} {Quantum erasers and probing classifications of entanglement via
  nuclear magnetic resonance},\ }\href@noop {} {\bibfield  {journal} {\bibinfo
  {journal} {Phys. Rev. A}\ }\textbf {\bibinfo {volume} {66}},\ \bibinfo
  {pages} {012309} (\bibinfo {year} {2002})}\BibitemShut {NoStop}%
\bibitem [{\citenamefont {Peng}\ \emph {et~al.}(2003)\citenamefont {Peng},
  \citenamefont {Zhu}, \citenamefont {Fang}, \citenamefont {Feng},
  \citenamefont {Liu},\ and\ \citenamefont {Gao}}]{peng2003interferometric}%
  \BibitemOpen
  \bibfield  {author} {\bibinfo {author} {\bibfnamefont {X.}~\bibnamefont
  {Peng}}, \bibinfo {author} {\bibfnamefont {X.}~\bibnamefont {Zhu}}, \bibinfo
  {author} {\bibfnamefont {X.}~\bibnamefont {Fang}}, \bibinfo {author}
  {\bibfnamefont {M.}~\bibnamefont {Feng}}, \bibinfo {author} {\bibfnamefont
  {M.}~\bibnamefont {Liu}},\ and\ \bibinfo {author} {\bibfnamefont
  {K.}~\bibnamefont {Gao}},\ }\bibfield  {title} {\bibinfo {title} {An
  interferometric complementarity experiment in a bulk nuclear magnetic
  resonance ensemble},\ }\href@noop {} {\bibfield  {journal} {\bibinfo
  {journal} {Journal of Physics A: Mathematical and General}\ }\textbf
  {\bibinfo {volume} {36}},\ \bibinfo {pages} {2555} (\bibinfo {year}
  {2003})}\BibitemShut {NoStop}%
\bibitem [{\citenamefont {Zheng}\ \emph {et~al.}(2015)\citenamefont {Zheng},
  \citenamefont {Zhong}, \citenamefont {Xu}, \citenamefont {Wang},
  \citenamefont {Wang}, \citenamefont {Shen}, \citenamefont {Yang},
  \citenamefont {Martinis}, \citenamefont {Cleland},\ and\ \citenamefont
  {Han}}]{zheng2015quantum}%
  \BibitemOpen
  \bibfield  {author} {\bibinfo {author} {\bibfnamefont {S.-B.}\ \bibnamefont
  {Zheng}}, \bibinfo {author} {\bibfnamefont {Y.-P.}\ \bibnamefont {Zhong}},
  \bibinfo {author} {\bibfnamefont {K.}~\bibnamefont {Xu}}, \bibinfo {author}
  {\bibfnamefont {Q.-J.}\ \bibnamefont {Wang}}, \bibinfo {author}
  {\bibfnamefont {H.}~\bibnamefont {Wang}}, \bibinfo {author} {\bibfnamefont
  {L.-T.}\ \bibnamefont {Shen}}, \bibinfo {author} {\bibfnamefont {C.-P.}\
  \bibnamefont {Yang}}, \bibinfo {author} {\bibfnamefont {J.~M.}\ \bibnamefont
  {Martinis}}, \bibinfo {author} {\bibfnamefont {A.~N.}\ \bibnamefont
  {Cleland}},\ and\ \bibinfo {author} {\bibfnamefont {S.-Y.}\ \bibnamefont
  {Han}},\ }\bibfield  {title} {\bibinfo {title} {Quantum delayed-choice
  experiment with a beam splitter in a quantum superposition},\ }\href@noop {}
  {\bibfield  {journal} {\bibinfo  {journal} {Phys. Rev. Lett.}\ }\textbf
  {\bibinfo {volume} {115}},\ \bibinfo {pages} {260403} (\bibinfo {year}
  {2015})}\BibitemShut {NoStop}%
\bibitem [{\citenamefont {Liu}\ \emph {et~al.}(2017)\citenamefont {Liu},
  \citenamefont {Xu}, \citenamefont {Wang}, \citenamefont {Zheng},
  \citenamefont {Roy}, \citenamefont {Kundu}, \citenamefont {Chand},
  \citenamefont {Ranadive}, \citenamefont {Vijay}, \citenamefont {Song},
  \citenamefont {Duan},\ and\ \citenamefont {Sun}}]{liu2017atwofold}%
  \BibitemOpen
  \bibfield  {author} {\bibinfo {author} {\bibfnamefont {K.}~\bibnamefont
  {Liu}}, \bibinfo {author} {\bibfnamefont {Y.}~\bibnamefont {Xu}}, \bibinfo
  {author} {\bibfnamefont {W.}~\bibnamefont {Wang}}, \bibinfo {author}
  {\bibfnamefont {S.-B.}\ \bibnamefont {Zheng}}, \bibinfo {author}
  {\bibfnamefont {T.}~\bibnamefont {Roy}}, \bibinfo {author} {\bibfnamefont
  {S.}~\bibnamefont {Kundu}}, \bibinfo {author} {\bibfnamefont
  {M.}~\bibnamefont {Chand}}, \bibinfo {author} {\bibfnamefont
  {A.}~\bibnamefont {Ranadive}}, \bibinfo {author} {\bibfnamefont
  {R.}~\bibnamefont {Vijay}}, \bibinfo {author} {\bibfnamefont
  {Y.}~\bibnamefont {Song}}, \bibinfo {author} {\bibfnamefont {L.}~\bibnamefont
  {Duan}},\ and\ \bibinfo {author} {\bibfnamefont {L.}~\bibnamefont {Sun}},\
  }\bibfield  {title} {\bibinfo {title} {A twofold quantum delayed-choice
  experiment in a superconducting circuit},\ }\href@noop {} {\bibfield
  {journal} {\bibinfo  {journal} {Science Advances}\ }\textbf {\bibinfo
  {volume} {3}},\ \bibinfo {pages} {e1603159} (\bibinfo {year}
  {2017})}\BibitemShut {NoStop}%
\bibitem [{\citenamefont {Bienfait}\ \emph {et~al.}(2020)\citenamefont
  {Bienfait}, \citenamefont {Zhong}, \citenamefont {Chang}, \citenamefont
  {Chou}, \citenamefont {Conner}, \citenamefont {Dumur}, \citenamefont
  {Grebel}, \citenamefont {Peairs}, \citenamefont {Povey}, \citenamefont
  {Satzinger},\ and\ \citenamefont {Cleland}}]{bienfait2020quantum}%
  \BibitemOpen
  \bibfield  {author} {\bibinfo {author} {\bibfnamefont {A.}~\bibnamefont
  {Bienfait}}, \bibinfo {author} {\bibfnamefont {Y.~P.}\ \bibnamefont {Zhong}},
  \bibinfo {author} {\bibfnamefont {H.-S.}\ \bibnamefont {Chang}}, \bibinfo
  {author} {\bibfnamefont {M.-H.}\ \bibnamefont {Chou}}, \bibinfo {author}
  {\bibfnamefont {C.~R.}\ \bibnamefont {Conner}}, \bibinfo {author}
  {\bibfnamefont {E.}~\bibnamefont {Dumur}}, \bibinfo {author} {\bibfnamefont
  {J.}~\bibnamefont {Grebel}}, \bibinfo {author} {\bibfnamefont {G.~A.}\
  \bibnamefont {Peairs}}, \bibinfo {author} {\bibfnamefont {R.~G.}\
  \bibnamefont {Povey}}, \bibinfo {author} {\bibfnamefont {K.~J.}\ \bibnamefont
  {Satzinger}},\ and\ \bibinfo {author} {\bibfnamefont {A.~N.}\ \bibnamefont
  {Cleland}},\ }\bibfield  {title} {\bibinfo {title} {Quantum erasure using
  entangled surface acoustic phonons},\ }\href@noop {} {\bibfield  {journal}
  {\bibinfo  {journal} {Phys. Rev. X}\ }\textbf {\bibinfo {volume} {10}},\
  \bibinfo {pages} {021055} (\bibinfo {year} {2020})}\BibitemShut {NoStop}%
\bibitem [{\citenamefont {Kwiat}\ \emph {et~al.}(1992)\citenamefont {Kwiat},
  \citenamefont {Steinberg},\ and\ \citenamefont
  {Chiao}}]{kwiat1992observation}%
  \BibitemOpen
  \bibfield  {author} {\bibinfo {author} {\bibfnamefont {P.~G.}\ \bibnamefont
  {Kwiat}}, \bibinfo {author} {\bibfnamefont {A.~M.}\ \bibnamefont
  {Steinberg}},\ and\ \bibinfo {author} {\bibfnamefont {R.~Y.}\ \bibnamefont
  {Chiao}},\ }\bibfield  {title} {\bibinfo {title} {Observation of a ``quantum
  eraser'': A revival of coherence in a two-photon interference experiment},\
  }\href@noop {} {\bibfield  {journal} {\bibinfo  {journal} {Phys. Rev. A}\
  }\textbf {\bibinfo {volume} {45}},\ \bibinfo {pages} {7729} (\bibinfo {year}
  {1992})}\BibitemShut {NoStop}%
\bibitem [{\citenamefont {Schwindt}\ \emph {et~al.}(1999)\citenamefont
  {Schwindt}, \citenamefont {Kwiat},\ and\ \citenamefont
  {Englert}}]{schwindt1999quantitative}%
  \BibitemOpen
  \bibfield  {author} {\bibinfo {author} {\bibfnamefont {P.~D.~D.}\
  \bibnamefont {Schwindt}}, \bibinfo {author} {\bibfnamefont {P.~G.}\
  \bibnamefont {Kwiat}},\ and\ \bibinfo {author} {\bibfnamefont {B.-G.}\
  \bibnamefont {Englert}},\ }\bibfield  {title} {\bibinfo {title} {Quantitative
  wave-particle duality and nonerasing quantum erasure},\ }\href@noop {}
  {\bibfield  {journal} {\bibinfo  {journal} {Phys. Rev. A}\ }\textbf {\bibinfo
  {volume} {60}},\ \bibinfo {pages} {4285} (\bibinfo {year}
  {1999})}\BibitemShut {NoStop}%
\bibitem [{ibm(4 30)}]{ibmq}%
  \BibitemOpen
  \href@noop {} {\bibinfo {title} {{IBM} {Q}uantum}},\ \bibinfo {howpublished}
  {\url{https://quantum-computing.ibm.com}} (\bibinfo {year} {Access:
  2021-04-30})\BibitemShut {NoStop}%
\bibitem [{\citenamefont {Amico}\ and\ \citenamefont
  {Dittel}(2020)}]{amico2020simulation}%
  \BibitemOpen
  \bibfield  {author} {\bibinfo {author} {\bibfnamefont {M.}~\bibnamefont
  {Amico}}\ and\ \bibinfo {author} {\bibfnamefont {C.}~\bibnamefont {Dittel}},\
  }\bibfield  {title} {\bibinfo {title} {Simulation of wave-particle duality in
  multipath interferometers on a quantum computer},\ }\href@noop {} {\bibfield
  {journal} {\bibinfo  {journal} {Phys. Rev. A}\ }\textbf {\bibinfo {volume}
  {102}},\ \bibinfo {pages} {032605} (\bibinfo {year} {2020})}\BibitemShut
  {NoStop}%
\bibitem [{\citenamefont {Schwaller}\ \emph {et~al.}(2021)\citenamefont
  {Schwaller}, \citenamefont {Dupertuis},\ and\ \citenamefont
  {Javerzac-Galy}}]{schwaller2021evidence}%
  \BibitemOpen
  \bibfield  {author} {\bibinfo {author} {\bibfnamefont {N.}~\bibnamefont
  {Schwaller}}, \bibinfo {author} {\bibfnamefont {M.-A.}\ \bibnamefont
  {Dupertuis}},\ and\ \bibinfo {author} {\bibfnamefont {C.}~\bibnamefont
  {Javerzac-Galy}},\ }\bibfield  {title} {\bibinfo {title} {Evidence of the
  entanglement constraint on wave-particle duality using the {IBM Q} quantum
  computer},\ }\href@noop {} {\bibfield  {journal} {\bibinfo  {journal} {Phys.
  Rev. A}\ }\textbf {\bibinfo {volume} {103}},\ \bibinfo {pages} {022409}
  (\bibinfo {year} {2021})}\BibitemShut {NoStop}%
\bibitem [{\citenamefont {Pozzobom}\ \emph {et~al.}(2021)\citenamefont
  {Pozzobom}, \citenamefont {Basso},\ and\ \citenamefont
  {Maziero}}]{pozzobom2021experimental}%
  \BibitemOpen
  \bibfield  {author} {\bibinfo {author} {\bibfnamefont {M.~B.}\ \bibnamefont
  {Pozzobom}}, \bibinfo {author} {\bibfnamefont {M.~L.~W.}\ \bibnamefont
  {Basso}},\ and\ \bibinfo {author} {\bibfnamefont {J.}~\bibnamefont
  {Maziero}},\ }\bibfield  {title} {\bibinfo {title} {Experimental tests of the
  density matrix's property-based complementarity relations},\ }\href@noop {}
  {\bibfield  {journal} {\bibinfo  {journal} {Phys. Rev. A}\ }\textbf {\bibinfo
  {volume} {103}},\ \bibinfo {pages} {022212} (\bibinfo {year}
  {2021})}\BibitemShut {NoStop}%
\bibitem [{\citenamefont {Cleve}\ \emph {et~al.}(1998)\citenamefont {Cleve},
  \citenamefont {Ekert}, \citenamefont {Macchiavello},\ and\ \citenamefont
  {Mosca}}]{cleve1998quantum}%
  \BibitemOpen
  \bibfield  {author} {\bibinfo {author} {\bibfnamefont {R.}~\bibnamefont
  {Cleve}}, \bibinfo {author} {\bibfnamefont {A.}~\bibnamefont {Ekert}},
  \bibinfo {author} {\bibfnamefont {C.}~\bibnamefont {Macchiavello}},\ and\
  \bibinfo {author} {\bibfnamefont {M.}~\bibnamefont {Mosca}},\ }\bibfield
  {title} {\bibinfo {title} {Quantum algorithms revisited},\ }\href@noop {}
  {\bibfield  {journal} {\bibinfo  {journal} {Proceedings of the Royal Society
  of London. Series A: Mathematical, Physical and Engineering Sciences}\
  }\textbf {\bibinfo {volume} {454}},\ \bibinfo {pages} {339} (\bibinfo {year}
  {1998})}\BibitemShut {NoStop}%
\bibitem [{\citenamefont {Helstrom}(1976)}]{helstrom1976quantum}%
  \BibitemOpen
  \bibfield  {author} {\bibinfo {author} {\bibfnamefont {C.~W.}\ \bibnamefont
  {Helstrom}},\ }\href@noop {} {\emph {\bibinfo {title} {Quantum detection and
  estimation theory}}}\ (\bibinfo  {publisher} {Academic press},\ \bibinfo
  {year} {1976})\BibitemShut {NoStop}%
\bibitem [{\citenamefont {Scully}\ \emph {et~al.}(1991)\citenamefont {Scully},
  \citenamefont {Englert},\ and\ \citenamefont {Walther}}]{scully1991quantum}%
  \BibitemOpen
  \bibfield  {author} {\bibinfo {author} {\bibfnamefont {M.~O.}\ \bibnamefont
  {Scully}}, \bibinfo {author} {\bibfnamefont {B.-G.}\ \bibnamefont
  {Englert}},\ and\ \bibinfo {author} {\bibfnamefont {H.}~\bibnamefont
  {Walther}},\ }\bibfield  {title} {\bibinfo {title} {Quantum optical tests of
  complementarity},\ }\href@noop {} {\bibfield  {journal} {\bibinfo  {journal}
  {Nature}\ }\textbf {\bibinfo {volume} {351}},\ \bibinfo {pages} {111}
  (\bibinfo {year} {1991})}\BibitemShut {NoStop}%
\bibitem [{\citenamefont {Ionicioiu}\ and\ \citenamefont
  {Terno}(2011)}]{ionicioiu2011proposal}%
  \BibitemOpen
  \bibfield  {author} {\bibinfo {author} {\bibfnamefont {R.}~\bibnamefont
  {Ionicioiu}}\ and\ \bibinfo {author} {\bibfnamefont {D.~R.}\ \bibnamefont
  {Terno}},\ }\bibfield  {title} {\bibinfo {title} {Proposal for a quantum
  delayed-choice experiment},\ }\href@noop {} {\bibfield  {journal} {\bibinfo
  {journal} {Phys. Rev. Lett.}\ }\textbf {\bibinfo {volume} {107}},\ \bibinfo
  {pages} {230406} (\bibinfo {year} {2011})}\BibitemShut {NoStop}%
\bibitem [{qis(4 30)}]{qiskit}%
  \BibitemOpen
  \href@noop {} {\bibinfo {title} {Qiskit: An open-source {SDK} for working
  with quantum computers at the level of pulses, circuits, and algorithms}},\
  \bibinfo {howpublished} {\url{https://qiskit.org}} (\bibinfo {year} {Access:
  2021-04-30})\BibitemShut {NoStop}%
\bibitem [{\citenamefont {Chamberland}\ \emph {et~al.}(2020)\citenamefont
  {Chamberland}, \citenamefont {Zhu}, \citenamefont {Yoder}, \citenamefont
  {Hertzberg},\ and\ \citenamefont {Cross}}]{chamberland2020topological}%
  \BibitemOpen
  \bibfield  {author} {\bibinfo {author} {\bibfnamefont {C.}~\bibnamefont
  {Chamberland}}, \bibinfo {author} {\bibfnamefont {G.}~\bibnamefont {Zhu}},
  \bibinfo {author} {\bibfnamefont {T.~J.}\ \bibnamefont {Yoder}}, \bibinfo
  {author} {\bibfnamefont {J.~B.}\ \bibnamefont {Hertzberg}},\ and\ \bibinfo
  {author} {\bibfnamefont {A.~W.}\ \bibnamefont {Cross}},\ }\bibfield  {title}
  {\bibinfo {title} {Topological and subsystem codes on low-degree graphs with
  flag qubits},\ }\href@noop {} {\bibfield  {journal} {\bibinfo  {journal}
  {Phys. Rev. X}\ }\textbf {\bibinfo {volume} {10}},\ \bibinfo {pages} {011022}
  (\bibinfo {year} {2020})}\BibitemShut {NoStop}%
\bibitem [{\citenamefont {Blais}\ \emph {et~al.}(2007)\citenamefont {Blais},
  \citenamefont {Gambetta}, \citenamefont {Wallraff}, \citenamefont {Schuster},
  \citenamefont {Girvin}, \citenamefont {Devoret},\ and\ \citenamefont
  {Schoelkopf}}]{blais2007quantum}%
  \BibitemOpen
  \bibfield  {author} {\bibinfo {author} {\bibfnamefont {A.}~\bibnamefont
  {Blais}}, \bibinfo {author} {\bibfnamefont {J.}~\bibnamefont {Gambetta}},
  \bibinfo {author} {\bibfnamefont {A.}~\bibnamefont {Wallraff}}, \bibinfo
  {author} {\bibfnamefont {D.~I.}\ \bibnamefont {Schuster}}, \bibinfo {author}
  {\bibfnamefont {S.~M.}\ \bibnamefont {Girvin}}, \bibinfo {author}
  {\bibfnamefont {M.~H.}\ \bibnamefont {Devoret}},\ and\ \bibinfo {author}
  {\bibfnamefont {R.~J.}\ \bibnamefont {Schoelkopf}},\ }\bibfield  {title}
  {\bibinfo {title} {Quantum-information processing with circuit quantum
  electrodynamics},\ }\href@noop {} {\bibfield  {journal} {\bibinfo  {journal}
  {Physical Review A}\ }\textbf {\bibinfo {volume} {75}},\ \bibinfo {pages}
  {032329} (\bibinfo {year} {2007})}\BibitemShut {NoStop}%
\bibitem [{\citenamefont {Koch}\ \emph {et~al.}(2007)\citenamefont {Koch},
  \citenamefont {Yu}, \citenamefont {Gambetta}, \citenamefont {Houck},
  \citenamefont {Schuster}, \citenamefont {Majer}, \citenamefont {Blais},
  \citenamefont {Devoret}, \citenamefont {Girvin},\ and\ \citenamefont
  {Schoelkopf}}]{koch2007charge}%
  \BibitemOpen
  \bibfield  {author} {\bibinfo {author} {\bibfnamefont {J.}~\bibnamefont
  {Koch}}, \bibinfo {author} {\bibfnamefont {T.~M.}\ \bibnamefont {Yu}},
  \bibinfo {author} {\bibfnamefont {J.}~\bibnamefont {Gambetta}}, \bibinfo
  {author} {\bibfnamefont {A.~A.}\ \bibnamefont {Houck}}, \bibinfo {author}
  {\bibfnamefont {D.~I.}\ \bibnamefont {Schuster}}, \bibinfo {author}
  {\bibfnamefont {J.}~\bibnamefont {Majer}}, \bibinfo {author} {\bibfnamefont
  {A.}~\bibnamefont {Blais}}, \bibinfo {author} {\bibfnamefont {M.~H.}\
  \bibnamefont {Devoret}}, \bibinfo {author} {\bibfnamefont {S.~M.}\
  \bibnamefont {Girvin}},\ and\ \bibinfo {author} {\bibfnamefont {R.~J.}\
  \bibnamefont {Schoelkopf}},\ }\bibfield  {title} {\bibinfo {title}
  {Charge-insensitive qubit design derived from the cooper pair box},\
  }\href@noop {} {\bibfield  {journal} {\bibinfo  {journal} {Phys. Rev. A}\
  }\textbf {\bibinfo {volume} {76}},\ \bibinfo {pages} {042319} (\bibinfo
  {year} {2007})}\BibitemShut {NoStop}%
\bibitem [{\citenamefont {Majer}\ \emph {et~al.}(2007)\citenamefont {Majer},
  \citenamefont {Chow}, \citenamefont {Gambetta}, \citenamefont {Koch},
  \citenamefont {Johnson}, \citenamefont {Schreier}, \citenamefont {Frunzio},
  \citenamefont {Schuster}, \citenamefont {Houck}, \citenamefont {Wallraff}
  \emph {et~al.}}]{majer2007coupling}%
  \BibitemOpen
  \bibfield  {author} {\bibinfo {author} {\bibfnamefont {J.}~\bibnamefont
  {Majer}}, \bibinfo {author} {\bibfnamefont {J.}~\bibnamefont {Chow}},
  \bibinfo {author} {\bibfnamefont {J.}~\bibnamefont {Gambetta}}, \bibinfo
  {author} {\bibfnamefont {J.}~\bibnamefont {Koch}}, \bibinfo {author}
  {\bibfnamefont {B.}~\bibnamefont {Johnson}}, \bibinfo {author} {\bibfnamefont
  {J.}~\bibnamefont {Schreier}}, \bibinfo {author} {\bibfnamefont
  {L.}~\bibnamefont {Frunzio}}, \bibinfo {author} {\bibfnamefont
  {D.}~\bibnamefont {Schuster}}, \bibinfo {author} {\bibfnamefont {A.~A.}\
  \bibnamefont {Houck}}, \bibinfo {author} {\bibfnamefont {A.}~\bibnamefont
  {Wallraff}}, \emph {et~al.},\ }\bibfield  {title} {\bibinfo {title} {Coupling
  superconducting qubits via a cavity bus},\ }\href@noop {} {\bibfield
  {journal} {\bibinfo  {journal} {Nature}\ }\textbf {\bibinfo {volume} {449}},\
  \bibinfo {pages} {443} (\bibinfo {year} {2007})}\BibitemShut {NoStop}%
\bibitem [{\citenamefont {Krantz}\ \emph {et~al.}(2019)\citenamefont {Krantz},
  \citenamefont {Kjaergaard}, \citenamefont {Yan}, \citenamefont {Orlando},
  \citenamefont {Gustavsson},\ and\ \citenamefont
  {Oliver}}]{krantz2019aquantum}%
  \BibitemOpen
  \bibfield  {author} {\bibinfo {author} {\bibfnamefont {P.}~\bibnamefont
  {Krantz}}, \bibinfo {author} {\bibfnamefont {M.}~\bibnamefont {Kjaergaard}},
  \bibinfo {author} {\bibfnamefont {F.}~\bibnamefont {Yan}}, \bibinfo {author}
  {\bibfnamefont {T.~P.}\ \bibnamefont {Orlando}}, \bibinfo {author}
  {\bibfnamefont {S.}~\bibnamefont {Gustavsson}},\ and\ \bibinfo {author}
  {\bibfnamefont {W.~D.}\ \bibnamefont {Oliver}},\ }\bibfield  {title}
  {\bibinfo {title} {A quantum engineer's guide to superconducting qubits},\
  }\href@noop {} {\bibfield  {journal} {\bibinfo  {journal} {Applied Physics
  Reviews}\ }\textbf {\bibinfo {volume} {6}},\ \bibinfo {pages} {021318}
  (\bibinfo {year} {2019})}\BibitemShut {NoStop}%
\bibitem [{\citenamefont {Blais}\ \emph {et~al.}(2021)\citenamefont {Blais},
  \citenamefont {Grimsmo}, \citenamefont {Girvin},\ and\ \citenamefont
  {Wallraff}}]{blais2020circuit}%
  \BibitemOpen
  \bibfield  {author} {\bibinfo {author} {\bibfnamefont {A.}~\bibnamefont
  {Blais}}, \bibinfo {author} {\bibfnamefont {A.~L.}\ \bibnamefont {Grimsmo}},
  \bibinfo {author} {\bibfnamefont {S.~M.}\ \bibnamefont {Girvin}},\ and\
  \bibinfo {author} {\bibfnamefont {A.}~\bibnamefont {Wallraff}},\ }\bibfield
  {title} {\bibinfo {title} {Circuit quantum electrodynamics},\ }\href@noop {}
  {\bibfield  {journal} {\bibinfo  {journal} {Rev. Mod. Phys.}\ }\textbf
  {\bibinfo {volume} {93}},\ \bibinfo {pages} {025005} (\bibinfo {year}
  {2021})}\BibitemShut {NoStop}%
\bibitem [{\citenamefont {Alsina}\ and\ \citenamefont
  {Latorre}(2016)}]{alsina2016experimental}%
  \BibitemOpen
  \bibfield  {author} {\bibinfo {author} {\bibfnamefont {D.}~\bibnamefont
  {Alsina}}\ and\ \bibinfo {author} {\bibfnamefont {J.~I.}\ \bibnamefont
  {Latorre}},\ }\bibfield  {title} {\bibinfo {title} {Experimental test of
  {M}ermin inequalities on a five-qubit quantum computer},\ }\href@noop {}
  {\bibfield  {journal} {\bibinfo  {journal} {Physical Review A}\ }\textbf
  {\bibinfo {volume} {94}},\ \bibinfo {pages} {012314} (\bibinfo {year}
  {2016})}\BibitemShut {NoStop}%
\bibitem [{\citenamefont {Garc{\'\i}a-Mart{\'\i}n}\ and\ \citenamefont
  {Sierra}(2018)}]{garcia2018five}%
  \BibitemOpen
  \bibfield  {author} {\bibinfo {author} {\bibfnamefont {D.}~\bibnamefont
  {Garc{\'\i}a-Mart{\'\i}n}}\ and\ \bibinfo {author} {\bibfnamefont
  {G.}~\bibnamefont {Sierra}},\ }\bibfield  {title} {\bibinfo {title} {Five
  experimental tests on the 5-qubit {IBM} quantum computer},\ }\href@noop {}
  {\bibfield  {journal} {\bibinfo  {journal} {Journal of Applied Mathematics
  and Physics}\ }\textbf {\bibinfo {volume} {6}},\ \bibinfo {pages} {1460}
  (\bibinfo {year} {2018})}\BibitemShut {NoStop}%
\bibitem [{rig(4 30)}]{rigettiweb}%
  \BibitemOpen
  \href@noop {} {}\bibinfo {howpublished} {\url{https://www.rigetti.com}}
  (\bibinfo {year} {Access: 2021-04-30})\BibitemShut {NoStop}%
\bibitem [{goo(4 30)}]{googleqai}%
  \BibitemOpen
  \href@noop {} {}\bibinfo {howpublished} {\url{https://quantumai.google}}
  (\bibinfo {year} {Access: 2021-04-30})\BibitemShut {NoStop}%
\bibitem [{qua(4 30)}]{quantuminspire}%
  \BibitemOpen
  \href@noop {} {}\bibinfo {howpublished}
  {\url{https://www.quantum-inspire.com}} (\bibinfo {year} {Access:
  2021-04-30})\BibitemShut {NoStop}%
\bibitem [{\citenamefont {Chen}\ \emph {et~al.}(2019)\citenamefont {Chen},
  \citenamefont {Farahzad}, \citenamefont {Yoo},\ and\ \citenamefont
  {Wei}}]{chen2019detector}%
  \BibitemOpen
  \bibfield  {author} {\bibinfo {author} {\bibfnamefont {Y.}~\bibnamefont
  {Chen}}, \bibinfo {author} {\bibfnamefont {M.}~\bibnamefont {Farahzad}},
  \bibinfo {author} {\bibfnamefont {S.}~\bibnamefont {Yoo}},\ and\ \bibinfo
  {author} {\bibfnamefont {T.-C.}\ \bibnamefont {Wei}},\ }\bibfield  {title}
  {\bibinfo {title} {Detector tomography on {IBM} quantum computers and
  mitigation of an imperfect measurement},\ }\href@noop {} {\bibfield
  {journal} {\bibinfo  {journal} {Phys. Rev. A}\ }\textbf {\bibinfo {volume}
  {100}},\ \bibinfo {pages} {052315} (\bibinfo {year} {2019})}\BibitemShut
  {NoStop}%
\bibitem [{\citenamefont {Maciejewski}\ \emph {et~al.}(2020)\citenamefont
  {Maciejewski}, \citenamefont {Zimbor{\'{a}}s},\ and\ \citenamefont
  {Oszmaniec}}]{maciejewski2020mitigation}%
  \BibitemOpen
  \bibfield  {author} {\bibinfo {author} {\bibfnamefont {F.~B.}\ \bibnamefont
  {Maciejewski}}, \bibinfo {author} {\bibfnamefont {Z.}~\bibnamefont
  {Zimbor{\'{a}}s}},\ and\ \bibinfo {author} {\bibfnamefont {M.}~\bibnamefont
  {Oszmaniec}},\ }\bibfield  {title} {\bibinfo {title} {Mitigation of readout
  noise in near-term quantum devices by classical post-processing based on
  detector tomography},\ }\href@noop {} {\bibfield  {journal} {\bibinfo
  {journal} {{Quantum}}\ }\textbf {\bibinfo {volume} {4}},\ \bibinfo {pages}
  {257} (\bibinfo {year} {2020})}\BibitemShut {NoStop}%
\bibitem [{\citenamefont {Asfaw}\ \emph {et~al.}(2020)\citenamefont {Asfaw},
  \citenamefont {Bello}, \citenamefont {Ben-Haim}, \citenamefont {Bravyi},
  \citenamefont {Bronn}, \citenamefont {Capelluto}, \citenamefont {Vazquez},
  \citenamefont {Ceroni}, \citenamefont {Chen}, \citenamefont {Frisch},
  \citenamefont {Gambetta}, \citenamefont {Garion}, \citenamefont {Gil},
  \citenamefont {Gonzalez}, \citenamefont {Harkins}, \citenamefont {Imamichi},
  \citenamefont {McKay}, \citenamefont {Mezzacapo}, \citenamefont {Minev},
  \citenamefont {Movassagh}, \citenamefont {Nannicni}, \citenamefont {Nation},
  \citenamefont {Phan}, \citenamefont {Pistoia}, \citenamefont {Rattew},
  \citenamefont {Schaefer}, \citenamefont {Shabani}, \citenamefont {Smolin},
  \citenamefont {Temme}, \citenamefont {Tod}, \citenamefont {Wood},\ and\
  \citenamefont {Wootton}}]{qiskittextbook2020}%
  \BibitemOpen
  \bibfield  {author} {\bibinfo {author} {\bibfnamefont {A.}~\bibnamefont
  {Asfaw}}, \bibinfo {author} {\bibfnamefont {L.}~\bibnamefont {Bello}},
  \bibinfo {author} {\bibfnamefont {Y.}~\bibnamefont {Ben-Haim}}, \bibinfo
  {author} {\bibfnamefont {S.}~\bibnamefont {Bravyi}}, \bibinfo {author}
  {\bibfnamefont {N.}~\bibnamefont {Bronn}}, \bibinfo {author} {\bibfnamefont
  {L.}~\bibnamefont {Capelluto}}, \bibinfo {author} {\bibfnamefont {A.~C.}\
  \bibnamefont {Vazquez}}, \bibinfo {author} {\bibfnamefont {J.}~\bibnamefont
  {Ceroni}}, \bibinfo {author} {\bibfnamefont {R.}~\bibnamefont {Chen}},
  \bibinfo {author} {\bibfnamefont {A.}~\bibnamefont {Frisch}}, \bibinfo
  {author} {\bibfnamefont {J.}~\bibnamefont {Gambetta}}, \bibinfo {author}
  {\bibfnamefont {S.}~\bibnamefont {Garion}}, \bibinfo {author} {\bibfnamefont
  {L.}~\bibnamefont {Gil}}, \bibinfo {author} {\bibfnamefont {S.~D. L.~P.}\
  \bibnamefont {Gonzalez}}, \bibinfo {author} {\bibfnamefont {F.}~\bibnamefont
  {Harkins}}, \bibinfo {author} {\bibfnamefont {T.}~\bibnamefont {Imamichi}},
  \bibinfo {author} {\bibfnamefont {D.}~\bibnamefont {McKay}}, \bibinfo
  {author} {\bibfnamefont {A.}~\bibnamefont {Mezzacapo}}, \bibinfo {author}
  {\bibfnamefont {Z.}~\bibnamefont {Minev}}, \bibinfo {author} {\bibfnamefont
  {R.}~\bibnamefont {Movassagh}}, \bibinfo {author} {\bibfnamefont
  {G.}~\bibnamefont {Nannicni}}, \bibinfo {author} {\bibfnamefont
  {P.}~\bibnamefont {Nation}}, \bibinfo {author} {\bibfnamefont
  {A.}~\bibnamefont {Phan}}, \bibinfo {author} {\bibfnamefont {M.}~\bibnamefont
  {Pistoia}}, \bibinfo {author} {\bibfnamefont {A.}~\bibnamefont {Rattew}},
  \bibinfo {author} {\bibfnamefont {J.}~\bibnamefont {Schaefer}}, \bibinfo
  {author} {\bibfnamefont {J.}~\bibnamefont {Shabani}}, \bibinfo {author}
  {\bibfnamefont {J.}~\bibnamefont {Smolin}}, \bibinfo {author} {\bibfnamefont
  {K.}~\bibnamefont {Temme}}, \bibinfo {author} {\bibfnamefont
  {M.}~\bibnamefont {Tod}}, \bibinfo {author} {\bibfnamefont {S.}~\bibnamefont
  {Wood}},\ and\ \bibinfo {author} {\bibfnamefont {J.}~\bibnamefont
  {Wootton}},\ }\href@noop {} {\bibinfo {title} {Learn quantum computation
  using qiskit}},\ \bibinfo {howpublished}
  {\url{http://community.qiskit.org/textbook}} (\bibinfo {year}
  {2020})\BibitemShut {NoStop}%
\bibitem [{\citenamefont {Wald}\ and\ \citenamefont
  {Wolfowitz}(1940)}]{wald1940test}%
  \BibitemOpen
  \bibfield  {author} {\bibinfo {author} {\bibfnamefont {A.}~\bibnamefont
  {Wald}}\ and\ \bibinfo {author} {\bibfnamefont {J.}~\bibnamefont
  {Wolfowitz}},\ }\bibfield  {title} {\bibinfo {title} {On a test whether two
  samples are from the same population},\ }\href@noop {} {\bibfield  {journal}
  {\bibinfo  {journal} {The Annals of Mathematical Statistics}\ }\textbf
  {\bibinfo {volume} {11}},\ \bibinfo {pages} {147} (\bibinfo {year}
  {1940})}\BibitemShut {NoStop}%
\bibitem [{\citenamefont {Sundaresan}\ \emph {et~al.}(2020)\citenamefont
  {Sundaresan}, \citenamefont {Lauer}, \citenamefont {Pritchett}, \citenamefont
  {Magesan}, \citenamefont {Jurcevic},\ and\ \citenamefont
  {Gambetta}}]{sundaresan2020reducing}%
  \BibitemOpen
  \bibfield  {author} {\bibinfo {author} {\bibfnamefont {N.}~\bibnamefont
  {Sundaresan}}, \bibinfo {author} {\bibfnamefont {I.}~\bibnamefont {Lauer}},
  \bibinfo {author} {\bibfnamefont {E.}~\bibnamefont {Pritchett}}, \bibinfo
  {author} {\bibfnamefont {E.}~\bibnamefont {Magesan}}, \bibinfo {author}
  {\bibfnamefont {P.}~\bibnamefont {Jurcevic}},\ and\ \bibinfo {author}
  {\bibfnamefont {J.~M.}\ \bibnamefont {Gambetta}},\ }\bibfield  {title}
  {\bibinfo {title} {Reducing unitary and spectator errors in cross resonance
  with optimized rotary echoes},\ }\href@noop {} {\bibfield  {journal}
  {\bibinfo  {journal} {PRX Quantum}\ }\textbf {\bibinfo {volume} {1}},\
  \bibinfo {pages} {020318} (\bibinfo {year} {2020})}\BibitemShut {NoStop}%
\bibitem [{\citenamefont {Zhang}\ \emph
  {et~al.}(2003{\natexlab{a}})\citenamefont {Zhang}, \citenamefont {Vala},
  \citenamefont {Sastry},\ and\ \citenamefont {Whaley}}]{zhang2003geometric}%
  \BibitemOpen
  \bibfield  {author} {\bibinfo {author} {\bibfnamefont {J.}~\bibnamefont
  {Zhang}}, \bibinfo {author} {\bibfnamefont {J.}~\bibnamefont {Vala}},
  \bibinfo {author} {\bibfnamefont {S.}~\bibnamefont {Sastry}},\ and\ \bibinfo
  {author} {\bibfnamefont {K.~B.}\ \bibnamefont {Whaley}},\ }\bibfield  {title}
  {\bibinfo {title} {Geometric theory of nonlocal two-qubit operations},\
  }\href@noop {} {\bibfield  {journal} {\bibinfo  {journal} {Phys. Rev. A}\
  }\textbf {\bibinfo {volume} {67}},\ \bibinfo {pages} {042313} (\bibinfo
  {year} {2003}{\natexlab{a}})}\BibitemShut {NoStop}%
\bibitem [{\citenamefont {Bremner}\ \emph {et~al.}(2002)\citenamefont
  {Bremner}, \citenamefont {Dawson}, \citenamefont {Dodd}, \citenamefont
  {Gilchrist}, \citenamefont {Harrow}, \citenamefont {Mortimer}, \citenamefont
  {Nielsen},\ and\ \citenamefont {Osborne}}]{bremmer2002practical}%
  \BibitemOpen
  \bibfield  {author} {\bibinfo {author} {\bibfnamefont {M.~J.}\ \bibnamefont
  {Bremner}}, \bibinfo {author} {\bibfnamefont {C.~M.}\ \bibnamefont {Dawson}},
  \bibinfo {author} {\bibfnamefont {J.~L.}\ \bibnamefont {Dodd}}, \bibinfo
  {author} {\bibfnamefont {A.}~\bibnamefont {Gilchrist}}, \bibinfo {author}
  {\bibfnamefont {A.~W.}\ \bibnamefont {Harrow}}, \bibinfo {author}
  {\bibfnamefont {D.}~\bibnamefont {Mortimer}}, \bibinfo {author}
  {\bibfnamefont {M.~A.}\ \bibnamefont {Nielsen}},\ and\ \bibinfo {author}
  {\bibfnamefont {T.~J.}\ \bibnamefont {Osborne}},\ }\bibfield  {title}
  {\bibinfo {title} {Practical scheme for quantum computation with any
  two-qubit entangling gate},\ }\href@noop {} {\bibfield  {journal} {\bibinfo
  {journal} {Phys. Rev. Lett.}\ }\textbf {\bibinfo {volume} {89}},\ \bibinfo
  {pages} {247902} (\bibinfo {year} {2002})}\BibitemShut {NoStop}%
\bibitem [{\citenamefont {Zhang}\ \emph
  {et~al.}(2003{\natexlab{b}})\citenamefont {Zhang}, \citenamefont {Vala},
  \citenamefont {Sastry},\ and\ \citenamefont {Whaley}}]{zhang2003exact}%
  \BibitemOpen
  \bibfield  {author} {\bibinfo {author} {\bibfnamefont {J.}~\bibnamefont
  {Zhang}}, \bibinfo {author} {\bibfnamefont {J.}~\bibnamefont {Vala}},
  \bibinfo {author} {\bibfnamefont {S.}~\bibnamefont {Sastry}},\ and\ \bibinfo
  {author} {\bibfnamefont {K.~B.}\ \bibnamefont {Whaley}},\ }\bibfield  {title}
  {\bibinfo {title} {Exact two-qubit universal quantum circuit},\ }\href@noop
  {} {\bibfield  {journal} {\bibinfo  {journal} {Phys. Rev. Lett.}\ }\textbf
  {\bibinfo {volume} {91}},\ \bibinfo {pages} {027903} (\bibinfo {year}
  {2003}{\natexlab{b}})}\BibitemShut {NoStop}%
\bibitem [{\citenamefont {Dawson}\ and\ \citenamefont
  {Nielsen}(2006)}]{dawson2006solovaykitaev}%
  \BibitemOpen
  \bibfield  {author} {\bibinfo {author} {\bibfnamefont {C.~M.}\ \bibnamefont
  {Dawson}}\ and\ \bibinfo {author} {\bibfnamefont {M.~A.}\ \bibnamefont
  {Nielsen}},\ }\bibfield  {title} {\bibinfo {title} {The solovay-kitaev
  algorithm},\ }\href@noop {} {\bibfield  {journal} {\bibinfo  {journal}
  {Quantum Info. Comput.}\ }\textbf {\bibinfo {volume} {6}},\ \bibinfo {pages}
  {81–95} (\bibinfo {year} {2006})}\BibitemShut {NoStop}%
\bibitem [{\citenamefont {Wallman}\ and\ \citenamefont
  {Emerson}(2016)}]{wallman2016noise}%
  \BibitemOpen
  \bibfield  {author} {\bibinfo {author} {\bibfnamefont {J.~J.}\ \bibnamefont
  {Wallman}}\ and\ \bibinfo {author} {\bibfnamefont {J.}~\bibnamefont
  {Emerson}},\ }\bibfield  {title} {\bibinfo {title} {Noise tailoring for
  scalable quantum computation via randomized compiling},\ }\href
  {https://doi.org/10.1103/PhysRevA.94.052325} {\bibfield  {journal} {\bibinfo
  {journal} {Phys. Rev. A}\ }\textbf {\bibinfo {volume} {94}},\ \bibinfo
  {pages} {052325} (\bibinfo {year} {2016})}\BibitemShut {NoStop}%
\bibitem [{\citenamefont {Sheldon}\ \emph {et~al.}(2016)\citenamefont
  {Sheldon}, \citenamefont {Magesan}, \citenamefont {Chow},\ and\ \citenamefont
  {Gambetta}}]{sheldon2016procedure}%
  \BibitemOpen
  \bibfield  {author} {\bibinfo {author} {\bibfnamefont {S.}~\bibnamefont
  {Sheldon}}, \bibinfo {author} {\bibfnamefont {E.}~\bibnamefont {Magesan}},
  \bibinfo {author} {\bibfnamefont {J.~M.}\ \bibnamefont {Chow}},\ and\
  \bibinfo {author} {\bibfnamefont {J.~M.}\ \bibnamefont {Gambetta}},\
  }\bibfield  {title} {\bibinfo {title} {Procedure for systematically tuning up
  cross-talk in the cross-resonance gate},\ }\href
  {https://doi.org/10.1103/PhysRevA.93.060302} {\bibfield  {journal} {\bibinfo
  {journal} {Phys. Rev. A}\ }\textbf {\bibinfo {volume} {93}},\ \bibinfo
  {pages} {060302} (\bibinfo {year} {2016})}\BibitemShut {NoStop}%
\bibitem [{\citenamefont {Malekakhlagh}\ \emph {et~al.}(2020)\citenamefont
  {Malekakhlagh}, \citenamefont {Magesan},\ and\ \citenamefont
  {McKay}}]{malekakhlagh2020first}%
  \BibitemOpen
  \bibfield  {author} {\bibinfo {author} {\bibfnamefont {M.}~\bibnamefont
  {Malekakhlagh}}, \bibinfo {author} {\bibfnamefont {E.}~\bibnamefont
  {Magesan}},\ and\ \bibinfo {author} {\bibfnamefont {D.~C.}\ \bibnamefont
  {McKay}},\ }\bibfield  {title} {\bibinfo {title} {First-principles analysis
  of cross-resonance gate operation},\ }\href
  {https://doi.org/10.1103/PhysRevA.102.042605} {\bibfield  {journal} {\bibinfo
   {journal} {Phys. Rev. A}\ }\textbf {\bibinfo {volume} {102}},\ \bibinfo
  {pages} {042605} (\bibinfo {year} {2020})}\BibitemShut {NoStop}%
\end{thebibliography}%

\end{document}